# Nonlocal tensor order parameter of the deformed state of liquid crystals

A.N. Kudryavtsev[1,2,*], P.A. Purtov[1,3,†], and S.I. Trashkeev[1,4,‡]

[1]*Novosibirsk State University, Novosibirsk 630090, Russia*
[2]*Khristianovich Institute of Theoretical and Applied Mechanics, Novosibirsk 630090, Russia*
[3]*Voevodsky Institute of Chemical Kinetics and Combustion, Novosibirsk 630090, Russia*
[4]*Institute of Laser Physics, Novosibirsk 630090, Russia*

A generalized notion of a nonlocal tensor order parameter is introduced within the framework of the phenomenological approach. This parameter has the form of a traceless tensor correlation function or a tensor integral operator. Based on this form, the governing relations are written, which determine the steady states and phase transitions of the deformed liquid crystal. Linear relations for eigenfunctions of the introduced operator are derived. A principal drawback of currently available models of liquid crystals based on the local presentation of the tensor order parameter (equality of two Frank constants in the case of a quadratic form of the strain part of the free energy) is eliminated. Particular examples are considered, which demonstrate the model workability and the absence of contradictions in the model as well as its adequacy when describing small-scale structures.

## I. INTRODUCTION

Almost from the very beginning of theoretical investigations of liquid crystals (LCs), attempts were made to explain their properties on the basis of microscopic ideas on the structure of molecules composing them and on the nature of intermolecular forces acting between the molecules. Nevertheless, the most famous and most frequently used model is still the macroscopic continuum theory based on research of Oseen and Frank. This theory is based on the concept of a director, i.e., a unit vector **n**(**r**) whose direction at the point **r** coincides with the dominating direction of the major axes of molecules at the same point. For nonpolar molecules, only a certain axis is chosen rather than one of the opposite directions on this axis; therefore, the states with **n** and –**n** are physically equivalent. It is assumed that the free energy density associated with orientational elasticity of nematic and cholesteric LCs is a quadratic function of the director derivative with respect to the spatial coordinates:

$$F = F_0(\rho,T) + \frac{1}{2}K_1(\nabla\cdot\mathbf{n})^2 + \frac{1}{2}K_2(\mathbf{n}\cdot\nabla\times\mathbf{n}+q)^2 + \frac{1}{2}K_3(\mathbf{n}\times\nabla\times\mathbf{n})^2 + \frac{1}{2}(K_2+K_4)\nabla\cdot\left[(\mathbf{n}\cdot\nabla)\mathbf{n}-(\nabla\cdot\mathbf{n})\mathbf{n}\right]. \tag{1}$$

Here $F_0(\rho,T)$ is the free energy density in the nondeformed crystal, the next three terms correspond to splay (S), twist (T), and bend (B) deformations, the corresponding coefficients $K_{1\text{-}4}(\rho,T)$ are known as the elasticity moduli or Frank moduli, and $q$ is a parameter with an inverse length dimension, which defines the helix pitch distance of the cholesteric. The LC equilibrium equations can be obtained by calculating deviations from the free energy, i.e., deviations of the spatial integral from Eq. (1); in this case, the last term in Eq. (1) transforms to the surface integral and does not contribute to equations, but is taken into account in formulating the boundary conditions.

*alex@itam.nsc.ru
†purtov@kinetics.nsc.ru
‡sitrskv@mail.ru

Based on a detailed description, the hydrodynamic theory of liquid crystals was developed, mainly by Ericksen and Leslie. In the Ericksen-Leslie theory, the equations for usual hydrodynamic variables used to describe anisotropic fluid flows are supplemented with an equation that describes the time evolution of the director field. The basic notions and the results of the classical continuum theory can be found in many monographs and textbooks dealing with the LC topic (see, e.g., [1-3]).

The popularity of phenomenological concepts based on the notion of a director is caused by the fact that the theory formulated on the basis of these concepts allows one to describe a large class of flows observed in experiments. It is also a powerful tool for modeling the processes in various LC devices. Nevertheless, it was always clear that the area of applicability of this theory is restricted to problems where significant changes in the orientation direction occur at scale much greater than the radius of action of intermolecular forces.

Indeed, the description of the orientation order based on the director concept is neither most general nor sufficient for all situations. The orientation of an individual LC molecule possessing cylindrical symmetry can be completely defined by a unit vector $\boldsymbol{\xi}$ directed along the axis of symmetry of this molecule. The paired potential of interaction between the molecules can be then expressed as a function of traceless tensors of the second rank related to two molecules:

$$\hat{q} = \boldsymbol{\xi} \otimes \boldsymbol{\xi} - \frac{1}{3}\hat{I}. \tag{2}$$

Here $\otimes$ means a tensor product and the unit tensor is denoted by $\hat{I}$.

Calculating the thermal mean of Eq. (2) with the use of an appropriate distribution function of molecules over the angles, we obtain the so-called tensor order parameter, which was first introduced by de Gennes [4]:

$$\hat{Q} = Q\left(\mathbf{n} \otimes \mathbf{n} - \frac{1}{3}\hat{I}\right) + P\left(\mathbf{d} \otimes \mathbf{d} - \mathbf{h} \otimes \mathbf{h}\right). \tag{3}$$

Here $\mathbf{d}$ and $\mathbf{h}$ are two unit vectors orthogonal to the director $\mathbf{n}$ and to each other; $Q$ and $P$ are scalar parameters.

For molecules not possessing rotational symmetry, in addition to $\boldsymbol{\xi}$, one has to introduce two more unit vectors $\boldsymbol{\eta}$ and $\boldsymbol{\zeta}$, which are orthogonal to $\boldsymbol{\xi}$ and to each other and which are directed along two other major axes of the molecule. The interaction potential of such molecules can also include symmetric traceless tensors:

$$\hat{b} = \boldsymbol{\eta} \otimes \boldsymbol{\eta} - \boldsymbol{\zeta} \otimes \boldsymbol{\zeta}. \tag{4}$$

After averaging, we obtain one more tensor order parameter

$$\hat{B} = Q'\left(\mathbf{n} \otimes \mathbf{n} - \frac{1}{3}\hat{I}\right) + P'\left(\mathbf{d} \otimes \mathbf{d} - \mathbf{h} \otimes \mathbf{h}\right). \tag{5}$$

The conventional hydrodynamic Ericksen-Leslie theory describes the case where $P = Q' = P' = 0$, and $Q$ is assumed to be constant. A similar theory formulated in terms of two director fields can be also developed for biaxial nematic liquid crystals [5].

The weak point of the theory is the assumption about constant values of scalar order parameters, such as $Q$, $P$, $Q'$, and $P'$. Obviously, in order to describe the transition to an isotropic fluid, the order parameters should be at least temperature-dependent. Their nonconstancy can be essential in the vicinity of singular points and vector field lines where the director direction becomes uncertain. It is clear that the conventional hydrodynamic theory fails to describe the internal structure of linear defects (disclinations), leading to infinite values of the total disclination energy [2].

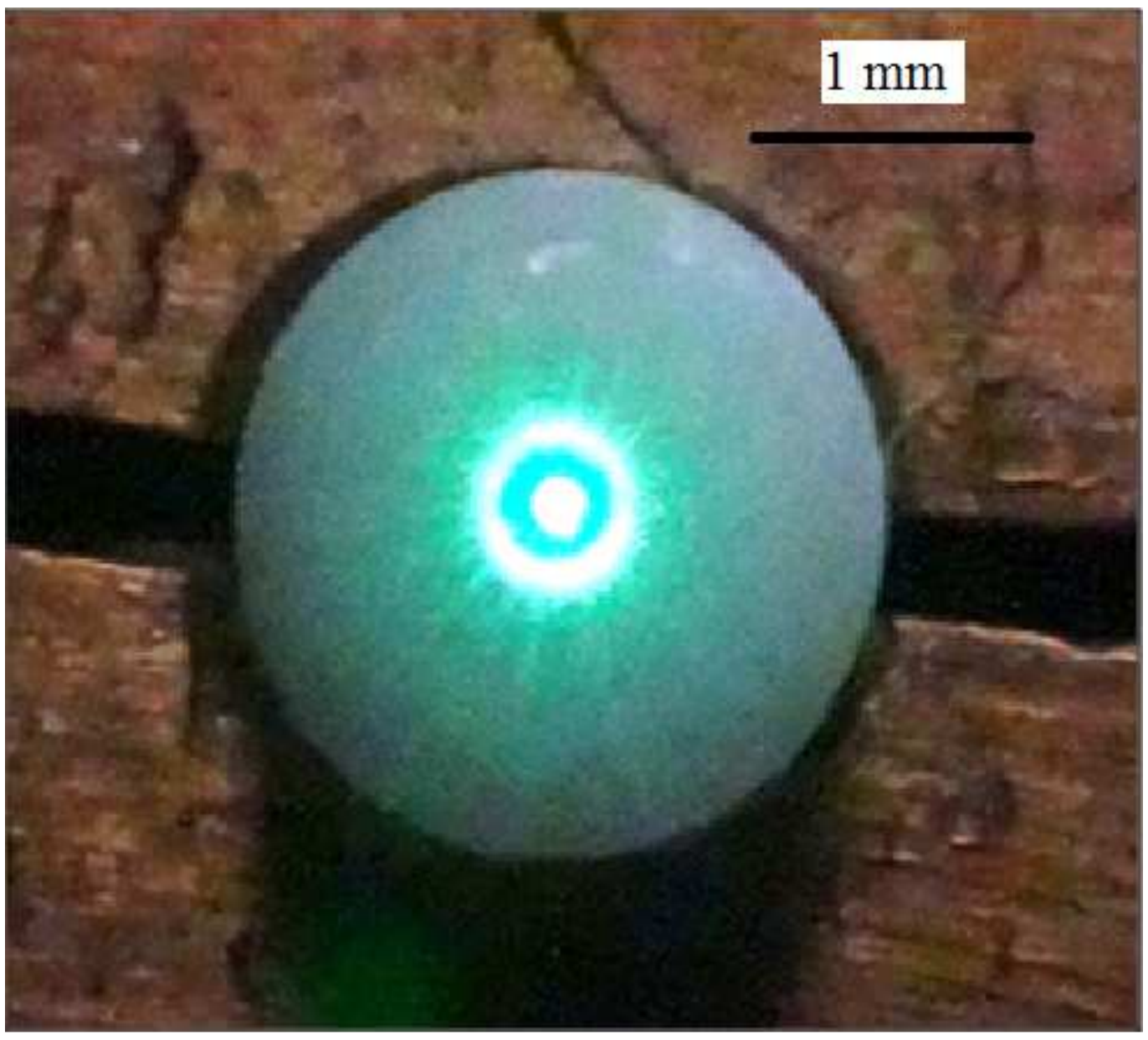

FIG. 1. Ring-shaped structure of the third harmonic generation zone in an NLC droplet at the end face of an optical fiber. The source is a femtosecond laser beam with a wavelength of 1560 nm; the mean power is ≈75 mW. The conversion efficiency is ≈ 15% [10].

Meanwhile, rapid recent development of new directions in LC research, which are primarily associated with nanotechnologies and nonlinear photonics, necessitated obtaining a correct description of phenomena that occur in liquid crystals at small spatial (< 10 μm) and temporal ($<10^{-4}$ s) scales. Such investigations, which are now gradually converted to applications, include activities in the field of composite media based on LC mixtures with nanoparticles and polymers and also those dealing with transformation and generation of optical radiation by liquid crystals [6, 7].

As an example, we can mention generation of harmonics (Fig. 1), difference frequencies, and spectral supercontinuum under conditions with high localization of laser radiation [7-11]. The use of fiber laser systems integrated with an LC medium provided high light energy concentrations (at sizes smaller than ~10 μm) at moderate mean power values (10 – 100 mW) of initial radiation [7, 9, 10]. Such concentrations provided a possibility of observations of numerous orientation-structural phenomena in NLCs, including such interesting effects as light-induced generation and destruction of disclinations, jumplike changes and hysteresis loops of the disclination force (Fig. 2), and also a strong dependence of the observed phenomena on the sample temperature (Fig. 3). Disclination origination may be initiated by a localized heat source generating a radial heat flux. The analysis of results reported in [7, 9, 10] allowed us to conclude that generation of diclinations and changes in their force can be considered as a phase transition with a complicated light-induced and temperature (due to nonlinear absorption) source. High-efficiency (up to several tens of percent of the input power) conversion of the radiation frequency is usually accompanied by generation of a disclination or by a change in the disclination force. The observed nonlinear optical and structural phenomena occur in a narrow ranges of radiation power and temperature (around the expected phase transitions) and are concentrated in microscopic (<10 μm) regions of LCs with the maximum strain or at the disclination center.

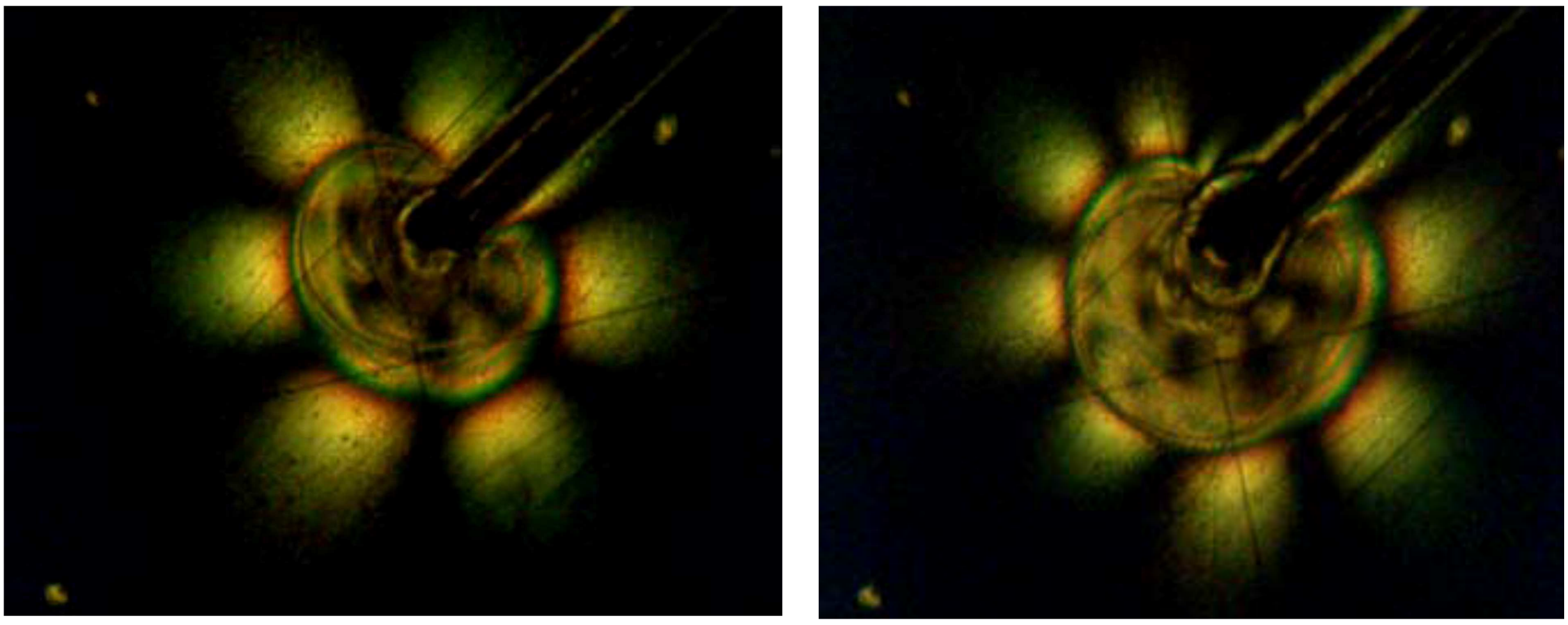

FIG.2. End face of an optical fiber ~100 μm in diameter embedded into an NLC layer (inserted between two glass substrates) in an external electrostatic field. Jumplike change in the disclination force *s*: *s* = 3/2 (left) and *s* = 2 (right). The source is a continuous laser beam with a wavelength of 1480 nm; the power is ≈120 mW [7, 9, 10].

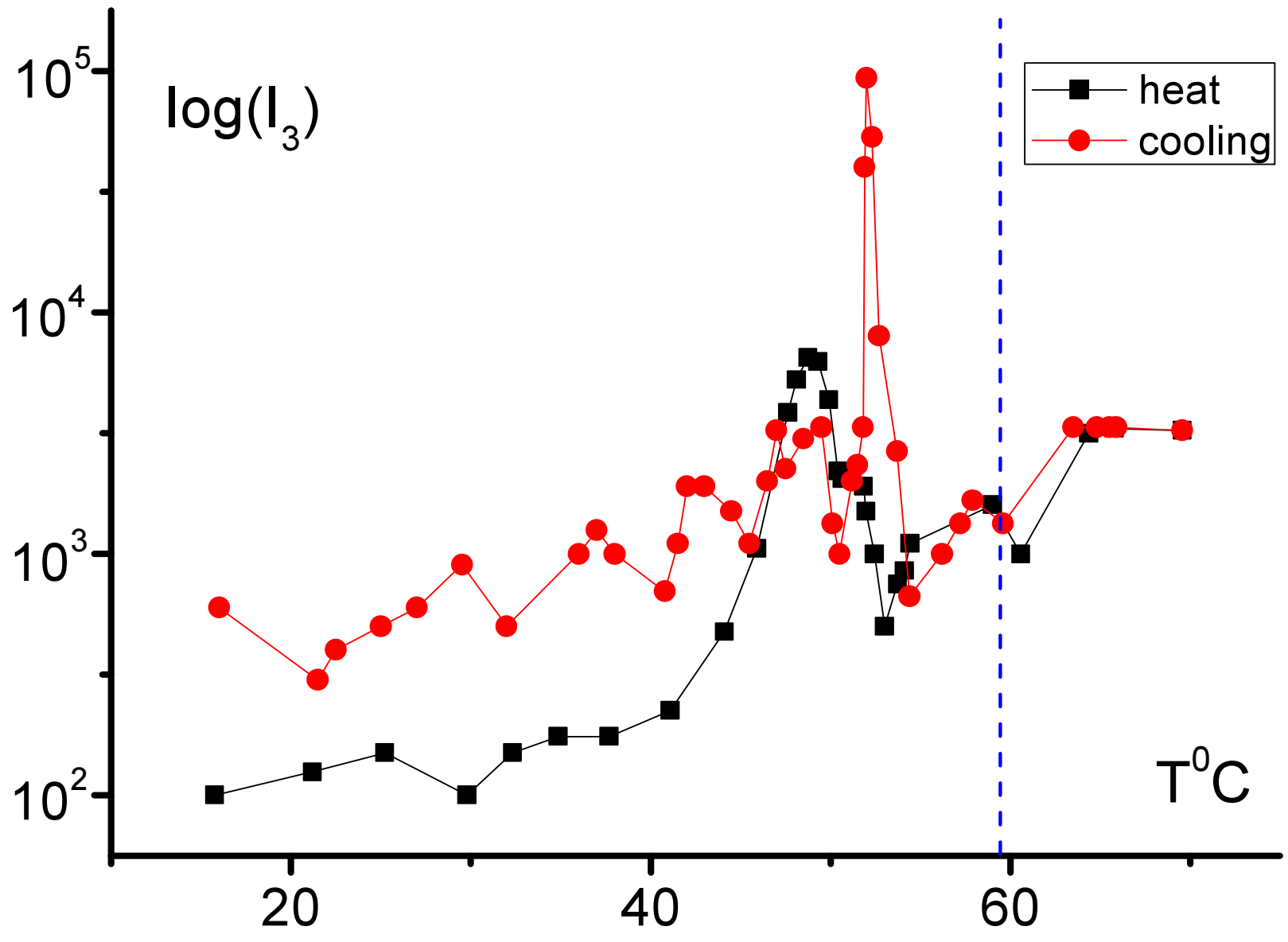

FIG. 3. Intensity of third harmonic generation versus temperature in an NLC droplet at the end gave of an optical fiber (Fig. 1 corresponds to peak generation). The hysteresis loop characterizes heating and subsequent cooling of the droplet. The blue dashed line shows the phase transition point $T_{NI}$≈59ºC [10].

The description of other light-induced phenomena in liquid crystals, e.g., the formation of solitons or the so-called nematicons (see the review in [12]), cannot be completely correct without due allowance for radiation absorption and for emergence of temperature gradients in the medium and, as a consequence, of heat fluxes. This is particularly true if LCs and their composites (with dyes or polymers) are affected by localized radiation [13, 14].

To confirm the influence of heat fluxes, orientation changes in NLCs without the laser action were studied. A temperature-induced effect on orientation not accompanied by the emergence of convective heat fluxes was observed in [15, 16]; it is impossible to consider localized phenomena, first of all, light-induced generation of disclinations, with this effect being ignored. Another effect of temperature on orientation that occurs due to flexoelectricity-type polarizability was considered in [17]; this effect should be also taken into account in the case of radiation absorption in LCs.

The first theory that took into account the change in the scalar order parameter $Q$ was proposed for uniaxial nematics in [18], where $Q$ was treated as a new variable determining the LC state together with $\mathbf{n}$. This theory did not become very popular. The point is that it is more natural to consider the effects associated with changes in orderliness on the basis of a theory whose principal variables that describe the orientation state are the components of the orientation tensor $\hat{Q}$ at $P = 0$ (3). Theories of this kind, which transform to the Ericksen-Leslie theory in the limit of a constant parameter $Q$, were formulated by Beris and Edwards [19] and by Qian and Sheng [20]. It was demonstrated by Sonnet et al. [21] that these two theories differ by terms in the dissipative function.

Numerical algorithms were developed for solving dynamic equations within the framework of both formalisms, and both theories were used for calculating particular phenomena of LC physics (see [22-24] and [25-26], respectively).

In addition to continuum approaches, which use the director vector field or the orientation tensor to describe the LC orientation state, much attention has been paid recently to approaches that take into account the molecular structure of the substance in this or that manner. A kinetic approach based on the nonlinear Smoluchowski equation (Fokker-Planck equation) for a nonequilibrium orientation distribution function evolving under the action of hydrodynamic, Brownian, and intermolecular forces was developed in [27, 28]. Computer simulation of liquid crystals based on a microscopic atomistic approach with the use of the Monte Carlo and molecular dynamics methods is being intensely developed. A review of publications on this topic can be found, e.g., in [25, 29]. In our opinion, the molecular dynamics method is of particular interest: in contrast to the Monte Carlo method, it can be used for simulating not only equilibrium properties of LCs, but also their dynamics.

A typical feature of the molecular dynamics method is its nonlocal character. The force acting on a molecule is determined by the positions and orientations of other molecules in a certain neighborhood around this molecule; a key element of the method is a potential with a small, but finite radius of its action, which describes interaction of molecules in the sense of the Newtonian action at a certain distance. The energy of interaction is then expressed by a double sum over molecule pairs.

It seems obvious that the theory that claims to provide a correct description of phenomena at mesoscopic (intermediate between molecular and macroscopic) scales has to include essential elements of both macroscopic continuum theories and microscopic description. In fact, such a theory must have a nonlocal character.

It should be noted that nonlocal theories are not new in continuum mechanics. Since the 1970s [30], such approaches have been developed within the framework of the elasticity theory to describe crack propagation, explain dispersion of elastic plane waves with high-frequencies, etc. Nonlocal theories naturally arise when discrete models of continuous media are considered [31]. A review of various nonlocal theories in continuum mechanics can be found in [32]. Formulating the governing equations of motion of continuous media in the form of integral equations become recently popular for modeling fracture of solids; this formulation was first proposed by Silling [33] and became known as peridynamics [34,35].

The possibility of nonlocal formulation was also considered for LC mechanics [36]; though this paper describes only the general conditions of including long-range interaction into equations that describe the dynamics of anisotropic fluids.

In the present paper, we propose a new phenomenological approach to the LC state analysis, which is based on a nonlocal description based on a tensor order parameter written for two points in space. This parameter is defined by a traceless tensor correlation function. The corresponding molecular analog of such an approach should be based on considering a two-particle kinetic distribution function. The classical theory proposed and developed by Oseen, Frank, and de Gennes, which is based on the concept of a local tensor order parameter and a single-particle distribution function, is a particular case (limiting approximation) of the proposed model. It is assumed that the proposed approach will provide a consistent description of physical phenomena that occur in liquid crystals at mesoscopic (intermediate between molecular and macroscopic) scales. In the present paper, we confine ourselves to considering only the orientation elasticity of liquid crystals and leave extension of the theory to phenomena including LC flows for the future.

## II. LOCAL PHENOMENOLOGICAL APPROACH

As it was already noted in Introduction, the most general method within the framework of the phenomenological approach for taking into account small-scale phenomena in liquid crystals is that proposed by de Gennes [4]. In this approach, steady-state LCs are considered with the use of the tensor order parameter or the orientation tensor $\hat{Q}$ (3). De Gennes [4] suggests that the free energy density of LCs should be considered in the form of an expansion in powers of $\hat{Q}=\left[Q_{ij}\right]$ and gradients $\nabla\hat{Q}=\left[\partial Q_{ij}/\partial x_k\right]=Q_{ij,k}$. In a later publication [37], this expansion was brought to the 4$^{\text{th}}$ power with squareness in terms of the gradient being smaller than $\hat{Q}\otimes\hat{Q}\otimes\left(\nabla\hat{Q}\otimes\nabla\hat{Q}\right)$. The high order of expansion is attributed to the necessity of consistency with experimental data; the primary task is to ensure different values of the Frank constants $K_1 \neq K_3$. The tensor order parameter can be presented in the form of Eq. (3), where $2Q/3$ and $-Q/3\pm P$ are eigenvalues, whereas **n**, **d,** and **h** are eigenvectors of tensor (3). In the general case, the LC description requires five independent values of the tensor $Q_{ij}$ ; if Eq. (3) is used, these may be $Q$, $P$, two components of the vector **n,** and one component of the vector **d**. For a uniaxial LC, we have $Q \neq 0$; the biaxial parameter is $P = 0$. The procedure of expansion of the free energy density in powers of the order parameter gives rise to unknown coefficients; moreover, the number of these coefficients becomes greater than 20 in the case of expansion of the NLC elastic energy to the 4$^{\text{th}}$ power, to the terms $\hat{Q}\otimes\hat{Q}\otimes\left(\nabla\hat{Q}\otimes\nabla\hat{Q}\right)$ [37]. The following question arises: How is it possible to determine these new coefficients if there are only four experimentally measured Frank constants in the uniaxial case? Spencer and Care [26] restricted the consideration to the cubic approximation by writing the bulk density of the LC free energy without external fields in the form

$$F=F_0+F_{el}+U, \tag{6}$$

where $F_0$ is the isotropic part independent of $Q_{ij}$, $F_{el}$ is the elastic part, $U$ is the Landau–de Gennes potential,

$$\begin{aligned}F_{el}&=\frac{1}{2}\left\{L_1Q_{ij,k}Q_{ij,k}+L_2Q_{ik,k}Q_{ij,j}+L_3Q_{ij,k}Q_{ik,j}+L_4Q_{ij}Q_{kl,i}Q_{kl,j}\right\}+\\&+2q\varepsilon_{ijk}\left(L_1Q_{il}Q_{kl,j}-L_4Q_{in}Q_{nm}Q_{km,j}\right)+\frac{3}{2}q^2\left(L_1Q_{ij}Q_{ji}-L_4Q_{ij}Q_{jk}Q_{ki}\right),\\U&=\frac{1}{2}\alpha Q_{ij}Q_{ji}-\beta Q_{ij}Q_{jk}Q_{ki}+\gamma Q_{ij}Q_{ji}Q_{kl}Q_{lk},\qquad i,j,k,l=1,2,3=x,y,z,\end{aligned} \tag{7}$$

$L_1$, $L_2$, $L_3$, and $L_4$ are the expansion coefficients (elastic parameters), $\varepsilon_{ijk}$ is the Levi-Civita symbol, $q = 2\pi/P_{ch}$, $P_{ch}$ is the pitch of chirality, and $\alpha$, $\beta$, and $\gamma$ are parameters responsible for phase transitions. Longa et al. [37] expanded $U$ with allowance for 6$^{\text{th}}$-order terms, $\sim (Sp\,\mathrm{Q}^2)^3$. Only four coefficients are left in the elastic energy equation (7) for them to be uniquely determined and for the inequality $K_1 \neq K_3$ to be satisfied. The total energy is expressed as an integral of the total free energy density $F$ over the volume, $F^{total}=\int FdV$, where $F$ is the sum determined from Eq. (6). The new constants were defined as follows. Dependence (3) with the LC uniaxiality condition $Q \neq 0$ ($-1/2 \leq Q \leq 1$), $P = 0$ was substituted into the expression for the elastic energy density $F_{el}$ in Eq. (7). Subsequent calculations yield the relations

$$F_{el} = \frac{1}{2}\left\{\frac{\partial Q}{\partial x_i} D_{ij} \frac{\partial Q}{\partial x_j} + P_k \frac{\partial Q^2}{\partial x_k}\right\} + F_{el}^{Frank}, \quad U = \frac{A}{2}Q^2 - BQ^3 + CQ^4, \tag{8}$$

where

$$D_{ij} = D_\perp \delta_{ij} + D_a n_i n_j, \quad D_a = D_\parallel - D_\perp,$$
$$\mathbf{P} = \mu_1 \mathbf{n}(\nabla \cdot \mathbf{n}) - \mu_3 \mathbf{n} \times \nabla \times \mathbf{n}, \tag{9}$$
$$F_{el}^{Frank} = \frac{K_1}{2}(\nabla \cdot \mathbf{n})^2 + \frac{K_2}{2}(\mathbf{n} \cdot \nabla \times \mathbf{n} + q)^2 + \frac{K_3}{2}(\mathbf{n} \times \nabla \times \mathbf{n})^2 + \frac{K_2 + K_4}{2}\left[\frac{\partial n_i}{\partial x_k}\frac{\partial n_k}{\partial x_i} - (\nabla \cdot \mathbf{n})^2\right].$$

The elastic Frank energy is identical to Eq. (1) with accuracy to the form of the last surface term. According to [26], the coefficients in Eqs. (8) and (9) are defined via $L_1$, $L_2$, $L_3$, and $L_4$ from Eq. (7) and $Q$ in the following way:

$$D_\parallel = \frac{4}{9}(L_2 + L_3 + L_4 Q), \; D_\perp = \frac{1}{9}(L_2 + L_3 - 2L_4 Q), \; D_a = D_\parallel - D_\perp = \frac{1}{3}(L_2 + L_3 + 2L_4 Q);$$
$$\mu_1 = \frac{2}{3}(L_2 - L_3), \quad \mu_3 = -\frac{\mu_1}{2} = -\frac{1}{3}(L_2 - L_3);$$
$$k_1 = \frac{K_1}{Q^2} = 2L_1 + L_2 + L_3 - \frac{2}{3}QL_4, \; k_2 = \frac{K_2}{Q^2} = 2L_1 - \frac{2}{3}QL_4, \; k_3 = \frac{K_3}{Q^2} = 2L_1 + L_2 + L_3 + \frac{4}{3}QL_4, \tag{10}$$
$$k_4 = \frac{K_4}{Q^2} = L_3 - 2L_1 + 2\frac{L_4}{3}Q, \; k_2 + k_4 = \frac{K_2 + K_4}{Q^2} = L_3, \; k_3 - k_1 = \frac{K_3 - K_1}{Q^2} = 2QL_4;$$
$$A = \frac{2}{3}\alpha, \; B = \frac{2}{9}\beta, \; C = \frac{4}{9}\gamma.$$

Here $A$, $B$, and $C$ are the coefficients of the Landau-de Gennes potential. At the same time, all coefficients, including the new coefficients (10), are defined via four Frank constants and $Q$ as

$$D_\parallel = \frac{2}{9Q^2}(K_1 - 2K_2 + K_3), \; D_\perp = \frac{1}{9Q^2}(2K_1 - K_2 - K_3), \; D_a = \frac{1}{3Q^2}(K_3 - K_2);$$
$$\mu_1 = \frac{2}{3Q^2}(K_1 - 3K_2 - 2K_4), \; \mu_3 = -\frac{1}{3Q^2}(K_1 - 3K_2 - 2K_4). \tag{11}$$

The coefficients (10), (11) were calculated taking into account the definition (3) unlike [26], where was assumed that $Q'_{ij} = Q'(n_i n_j - \delta_{ij}/3)/2$. The unique dependence of the resultant parameters on four known Frank constants introduced in the condition of a constant scalar order parameter is somewhat suspicious: now we have a new variational variable, which is independent of the director, whereas the number of the governing constants is unchanged. The equilibrium equations are determined on the basis of the variational principle ($F^{total}$ is the Lagrange function) according to the procedure

$$\frac{\partial}{\partial x_k}\frac{\partial F}{\partial(\partial n_i/\partial x_k)} - \frac{\partial F}{\partial n_i} = wn_i, \quad \frac{\partial}{\partial x_k}\frac{\partial F}{\partial(\partial Q/\partial x_k)} - \frac{\partial F}{\partial Q} = 0. \tag{12}$$

where $w$ is an undetermined Lagrange factor, which ensures the validity of the director uniqueness condition $\mathbf{n}^2 = 1$.

Relations (8) and (9) for the uniaxial case are fairly general because all calculations where the scalar order parameter $Q = Q(\mathbf{r})$ is introduced based on Eq. (3) yield the identical dependence (8) in the uniaxial case. These studies include [4, 20, 26, 37-39] and others where similar approaches are used, as well as the work of Ericksen [18], who introduced independent constants similar to $D$ and $\mu$. Obviously, the initial uniaxiality of LCs can be violated in the general case (the so-called "induced biaxiality") because of the emergence of two polar vectors whose directions do not necessarily coincide with the direction of **n**, $\mathbf{n}\times\mathbf{P}\sim\nabla\times\mathbf{n}\neq 0$, $\mathbf{n}\times\nabla Q\neq 0$. The form of the vector **P** is identical to the flexoelectric polarizability vector (9), and the gradient of the scalar order parameter $\nabla Q$ is something like an "orienting" field, which is similar to some extent to the electric or magnetic field. In this context, the coefficients $\mu_1$ and $\mu_3$ can be called "flex mechanical" or simply "flex coefficients". Here it is necessary to make a reservation. Formally, appearance of new vectors contradicts to the original assumption that the medium is uniaxial and is a consequence of approximation ensuing from the adopted condition $P=0$ in Eq. (3).

In [37], Eq. (10) is written via 22 introduced coefficients, $L_1$, $L_2$, …, $L_{22}$, and the general form of the Frank constants for any order $n+2$ of the free energy expansion in powers of $Q$ is given,

$$\frac{K_\alpha}{Q^2}=k_\alpha^{(0)}+k_\alpha^{(1)}Q+k_\alpha^{(2)}Q^2+\ \dots,\quad k_\alpha^{(n)}=\sum_m c_m^{(n)}L_m;\quad \alpha=1,2,3,4;\quad n,m=0,1,2,3,\ \dots,\tag{13}$$

where $c_\alpha^{(n)}$ are numerical factors. Expansions similar to Eq. (13) describe the coefficients $D$ and $\mu$ from Eq. (9). Based on Eq. (13), it is assumed that all necessary coefficients can be determined from experimental measurements of the Frank constants and their dependence on the order parameter, as was demonstrated, e.g., in [38]. This statement could be considered as valid if the scalar order parameter were independent of spatial variables. Moreover, at $K_\alpha=K_\alpha(Q(\mathbf{r}))$, it is not clear which variables are measured as elastic coefficients, e.g., in the case of the Freedericksz transition [1-3].

## III. INTEGRAL TENSOR ORDER PARAMETER

The definition of the order parameter $Q_{ij}$ introduced on the basis of Eq. (3) is local because the LC orderliness degree at the point **r** is determined by the state of orientation at the same point **r**. This statement is valid in a nondeformed (weakly deformed) LC that occupies a sufficiently large volume with distant boundaries. If this is not the case, it is necessary to take into account the influence of the neighboring points on the orientation order at the point **r**. Let us generalize Eq. (2) by considering the two-point tensor order parameter with a zero trace

$$S_{ij}(\mathbf{r},\mathbf{r}')=\ <\xi_i(\mathbf{r})\xi_j(\mathbf{r}')-\frac{\delta_{ij}}{3}\xi_k(\mathbf{r})\xi_k(\mathbf{r}')>,\tag{14}$$

where **r** and **r**' are the points located inside the volume $V$ occupied by the LC. Definition (14) has the form of a correlation tensor function. It is possible to use a slightly different definition in the form

$$\tilde{S}_{ij}(\mathbf{r},\mathbf{r}')=\ <\xi_i(\mathbf{r})\xi_j(\mathbf{r}')-\frac{\delta_{ij}}{3}>,\tag{15}$$

in this case, however, the zero trace should be understood as the following integral relation with the Kronecker delta and Dirac delta function:

$$\tilde{S}_{ii}(\mathbf{r},\mathbf{r})=\int_V S_{ij}(\mathbf{r},\mathbf{r}')\delta_{ij}\delta(\mathbf{r}-\mathbf{r}')dV'=0.$$

In contrast to the classical approach, these definitions require introduction of a two-point distribution function $\mathfrak{F}(\xi,\mathbf{r};\xi',\mathbf{r}')$, but this kinetic aspect is not considered at the moment. It should be noted that a similar statement can be found in [40]. Relations (14) and (15) have the symmetry $S_{ij}(\mathbf{r},\mathbf{r}')=S_{ji}(\mathbf{r}',\mathbf{r})$ and perform a transformation of vectors inside the volume $V$:

$$\mathbf{g}=\hat{S}\mathbf{f} \ \rightarrow \ g_i(\mathbf{r})=\int_V S_{ij}(\mathbf{r},\mathbf{r}')f_j(\mathbf{r}')\frac{dV'}{V}.$$

In what follows, integration is always assumed to be performed over the entire volume $V$ ($dV \equiv dxdydz$), and the kernel $S_{ij}$ is assumed to be real (Hermitian in the general case) and normalized, i.e., the following integral not equal to zero exists:

$$\iint \left|S_{ij}(\mathbf{r},\mathbf{r}')\right|^2 dVdV' \neq 0.$$

The usual order parameter is defined by the value at $\mathbf{r}$ = $\mathbf{r}'$ or by the convolution with the Dirac delta function $\delta(\mathbf{r}-\mathbf{r}')$,

$$Q_{ij}(\mathbf{r})=\int S_{ij}(\mathbf{r},\mathbf{r}')\delta(\mathbf{r}-\mathbf{r}')dV'=S_{ij}(\mathbf{r},\mathbf{r}).$$

The variable defined in Eq. (14) or (15) can be called the integral (or distributed, nonlocal, or two-point) tensor order parameter (ITOP). For a nondeformed (homogeneous in space, $\partial n_i/\partial x_k \rightarrow 0$) LC with sufficiently distance walls in the thermodynamic equilibrium state ($T, p \rightarrow const$), the following limiting condition is expected to be valid:

$$S_{ij}(\mathbf{r},\mathbf{r}') \underset{\substack{\partial n_i/\partial x_k=0,\\ T,p\,=const}}{\rightarrow} Q_0 V\delta(\mathbf{r}-\mathbf{r}')\left(n_i n_j-\frac{\delta_{ij}}{3}\right),$$

where $\mathbf{n}$ = $const$, $T$, $p$, and $-1/2 \leq Q_0 \leq 1$ are the equilibrium values of the director, temperature, pressure, and scalar order parameter. As in Eq. (6), the total free energy $F^{total}$ has the form of the sum of the energies of the isotropic state $F_0^{total}$, deformed LC $F_{el}^{total}$, and Landau-de Gennes potential $U^{total}$. To write the total free energy in terms of the ITOP, we need double integrals of the two-point energy densities over the volume:

$$F^{total}=F_0^{total}+F_{el}^{total}+U^{total},$$
$$F_0^{total}=\int F_0 dV,\quad F_{el}^{total}=\iint F_{el}\frac{dV'}{V}dV,\quad U^{total}=\iint U\frac{dV'}{V}dV. \tag{16}$$

In what follows, we omit terms with the meaning of "two-pointedness" for brevity and use the conventional terminology, assuming that it means the new definitions (14)-(16) or the abbreviation ITOP. Comments will be given where necessary.

The absence of symmetry only over the subscripts of the tensor operator $\hat{S}=S_{ij}(\mathbf{r},\mathbf{r}')\neq S_{ji}(\mathbf{r},\mathbf{r}')$ requires all permutations of the subscripts to be taken into account in writing the free energy density equation, namely,

$$F_{el}(\mathbf{r},\mathbf{r}')=\frac{1}{2}\left[\begin{array}{c} M_1S_{ij,k}S_{ij,k}+M_2S_{ik,k}S_{ij,j}+M_3S_{ij,k}S_{ik,j}+M_4S_{ji,k}S_{ij,k}+\\ +M_5S_{ki,k}S_{ij,j}+M_6S_{ji,k}S_{ik,j}+M_7S_{ki,k}S_{ji,j}+M_8S_{ji,k}S_{ki,j}+\\ +\hat{M}_q q\left(2\hat{\varepsilon}\otimes\hat{S}\otimes\nabla S+q\hat{S}\otimes\hat{S}\right)\end{array}\right]+\left[\frac{\partial}{\partial x_k}\rightarrow\frac{\partial}{\partial x'_k}\right],$$

$$S_{ij,k}=\frac{\partial S_{ij}(\mathbf{r},\mathbf{r}')}{\partial x_k},\qquad i,j,k,l=1,2,3=x,y,z, \tag{17}$$

$$U(\mathbf{r},\mathbf{r}')=\frac{\hat{\alpha}}{2}\hat{S}(\mathbf{r},\mathbf{r}')\otimes\hat{S}(\mathbf{r},\mathbf{r}')-\hat{\beta}\int\hat{S}(\mathbf{r},\mathbf{r}')\otimes\hat{S}(\mathbf{r},\mathbf{r}'')\otimes\hat{S}(\mathbf{r}',\mathbf{r}'')\frac{dV''}{V}+$$
$$+\hat{\gamma}\iint\hat{S}(\mathbf{r},\mathbf{r}')\otimes\hat{S}(\mathbf{r},\mathbf{r}'')\otimes\hat{S}(\mathbf{r}',\mathbf{r}''')\otimes\hat{S}(\mathbf{r}'',\mathbf{r}''')\frac{dV''dV'''}{V^2}.$$

In the elastic part of the free energy density $F_{el}$, the expression in the second square brackets $[\partial/\partial x_k \rightarrow \partial/\partial x'_k]$ means a term similar to that in the first square brackets, but differentiation is performed over the primed coordinates $\mathbf{r}'$. As can be easily seen, after double differentiation of Eq. (16), the terms in $[\partial/\partial x_k \rightarrow \partial/\partial x'_k]$ with differentiation with respect to $\mathbf{r}'$ yield the same result as the terms with differentiation with respect to **r** owing to the symmetry condition. Formally, by virtue of symmetry, it is sufficient to use one explicitly written term with differentiation with respect to **r**. In Eq. (17), it is sufficient to use only the quadratic forms of the ITOP gradient (14) or (15) for obtaining a noncontradictory description of the LC. The elastic part $F_{el}$ of Eq. (17) includes new constants $M_{1\ldots8}$ and a set of cholesteric constants $\hat{M}_q$, which are independent of the order parameter $S_{ij}$. In the cholesteric part and the function of the Landau-de Gennes potential $U(\mathbf{r},\mathbf{r}')$, the symbol ⊗ means all possible tensor products with different permutations of the tensor subscripts. For example, the first term with the coefficient $\hat{\alpha}/2$ has the form of a sum whose terms have different permutations of the subscripts:

$$\frac{\hat{\alpha}}{2}\int \hat{S}(\mathbf{r},\mathbf{r}')\otimes\hat{S}(\mathbf{r},\mathbf{r}')\frac{dV'}{V}+\ldots=\frac{1}{2}\int\left[\alpha_1 S_{ij}(\mathbf{r},\mathbf{r}')S_{ij}(\mathbf{r},\mathbf{r}')+\alpha_2 S_{ij}(\mathbf{r},\mathbf{r}')S_{ji}(\mathbf{r},\mathbf{r}')\right]\frac{dV'}{V}+\ldots .$$

The remaining terms denoted by the ellipsis, which are higher-order terms, are determined in a similar manner. The chiral term has the form

$$\hat{M}_q q\left(2\hat{\varepsilon}\otimes\hat{S}\otimes\nabla S+q\hat{S}\otimes\hat{S}\right)=$$

$$=2q\left(M_{1q}\varepsilon_{ijk}S_{il}S_{kl,j}+M_{2q}\varepsilon_{ijk}S_{il}S_{lk,j}+M_{3q}\varepsilon_{ijk}S_{li}S_{kl,j}+M_{4q}\varepsilon_{ijk}S_{li}S_{lk,j}\right)+q^2\left(M_{5q}S_{ij}S_{ij}+M_{6q}S_{ij}S_{ji}\right),$$

where $\hat{\varepsilon}=\left[\varepsilon_{ijk}\right]$ is the Levi-Civita operator. The sets of the parameters $\hat{\alpha}$, $\hat{\beta}$, $\hat{\gamma}$ in Eq. (17) are also independent of $S_{ij}$. Based on investigations of phase transitions in LCs [1-3, 6], it can be argued that their description requires at least three coefficients (one for each power of the order parameter) In the present ITOP variant, we can also use three coefficients (the number of coefficients can be increased if necessary) chosen in a manner to obtain the best possible fitting of experimental data. The total Landau-de Gennes potential $U^{total}$ can be written directly via the linearly independent ITOP invariants $I_\alpha$ whose number ($\alpha$ = 1,2, …) and form (for the power of four or smaller) will be determined later:

$$U^{total}=\int U\frac{dV'}{V}dV=U^{total}\left(I_1,I_2,\ \ldots\right). \qquad (18)$$

Formally, apart from individual derivatives with respect to the coordinates **r** and **r**', $F_{el}$ should also include mixed terms, such as

$$\frac{\partial S_{ij}(\mathbf{r},\mathbf{r}')}{\partial x_k}\frac{\partial S_{ij}(\mathbf{r},\mathbf{r}')}{\partial x'_k},\quad \frac{\partial S_{ik}(\mathbf{r},\mathbf{r}')}{\partial x_k}\frac{\partial S_{ij}(\mathbf{r},\mathbf{r}')}{\partial x'_j},\quad \frac{\partial S_{ki}(\mathbf{r},\mathbf{r}')}{\partial x_k}\frac{\partial S_{ij}(\mathbf{r},\mathbf{r}')}{\partial x'_j},\ \ldots .$$

Additives with cross derivatives do not contribute to volume deformation. However, this statement needs some disclaimer. Cross terms are mainly reduced to surface integrals and will be taken into account in imposing boundary conditions. An exception, which has to be considered in more detail, is the case of emerging boundaries with separation of phases, singular points, and other similar deformations. Such cases are not considered in the present paper; the reader is referred to the detailed discussion of the role of surface terms in [37] and references cited therein.

The equilibrium condition in the general form is derived from Eqs. (16) and (17) on the basis of variation as the equation in a six-dimensional space

$$\frac{\partial}{\partial x_k}\left(\frac{\partial F}{\partial S_{ij,k}}\right)+\frac{\partial}{\partial x'_k}\left(\frac{\partial F}{\partial S_{ij,k}}\right)-\frac{\partial F}{\partial S_{ij}}-\eta\delta_{ij}-\eta_{ij}=0, \qquad (19)$$

where $\eta$ and $\eta_{ij}$ are the Lagrangian multipliers introduced to satisfy the traceless condition and to ensure the symmetry of the tensor kernel. The scalar variable $\eta$ is calculated as follows: for definition (14)

$$\frac{\delta_{ij}}{3}\left[\frac{\partial}{\partial x_k}\left(\frac{\partial F}{\partial S_{ij,k}}\right)+\frac{\partial}{\partial x'_k}\left(\frac{\partial F}{\partial S_{ij,k}}\right)-\frac{\partial F}{\partial S_{ij}}\right]=\eta(\mathbf{r},\mathbf{r}'),$$

and, correspondingly, for definition (15)

$$\frac{\delta_{ij}}{3}\int\left[\frac{\partial}{\partial x_k}\left(\frac{\partial F}{\partial \tilde{S}_{ij,k}}\right)+\frac{\partial}{\partial x'_k}\left(\frac{\partial F}{\partial \tilde{S}_{ij,k}}\right)-\frac{\partial F}{\partial \tilde{S}_{ij}}\right]\delta(\mathbf{r}-\mathbf{r}')dV'=\eta(\mathbf{r},\mathbf{r}')\big|_{\mathbf{r}'=\mathbf{r}}\equiv\eta(\mathbf{r}).$$

The tensor parameter $\eta_{ij}$ has an antisymmetric form composed of a certain vector $\eta_l(\mathbf{r},\mathbf{r}')$ with a permutation of arguments

$$\eta_{ij}(\mathbf{r},\mathbf{r}')=\begin{pmatrix} 0 & \eta_z(\mathbf{r},\mathbf{r}') & -\eta_y(\mathbf{r},\mathbf{r}') \\ -\eta_z(\mathbf{r}',\mathbf{r}) & 0 & \eta_x(\mathbf{r},\mathbf{r}') \\ \eta_y(\mathbf{r}',\mathbf{r}) & -\eta_x(\mathbf{r}',\mathbf{r}) & 0 \end{pmatrix},$$

and is calculated by the relations

$$\eta_l=\frac{\varepsilon_{ijl}}{2}\left[\frac{\partial}{\partial x_k}\left(\frac{\partial F}{\partial \overleftrightarrow{S}_{ij,k}}\right)+\frac{\partial}{\partial x'_k}\left(\frac{\partial F}{\partial \overleftrightarrow{S}_{ij,k}}\right)-\frac{\partial F}{\partial \overleftrightarrow{S}_{ij}}\right],\quad \overleftrightarrow{S}_{ij}=\begin{cases} S_{ij}(\mathbf{r},\mathbf{r}')\ or\ \tilde{S}_{ij}(\mathbf{r},\mathbf{r}'), & i\ge j,\\ S_{ij}(\mathbf{r}',\mathbf{r})\ or\ \tilde{S}_{ij}(\mathbf{r}',\mathbf{r}), & i<j.\end{cases}$$

In the general case, the variational equations for the Landau-de Gennes potential (17) more than quadratic in terms of $S_{ij}$ have a nonlinear integrodifferential form.

Equation (19) can be written in a different form by using such charateristics of the integral operator as its eigenvalues and eigenvectors. Thus, for a real symmetric kernel $S_{ij}(\mathbf{r},\mathbf{r}')$, the relation

$$\lambda f_i(\mathbf{r})=\int S_{ij}(\mathbf{r},\mathbf{r}')f_j(\mathbf{r}')\frac{dV'}{V},$$

yields an infinite set of real triples $\lambda$ and $\mathbf{f}=(f_x,f_y,f_z)$, which can be conveniently written in the form of the sequences

$$\lambda\to\left(\lambda_f^{(\sigma)}\ \ \lambda_g^{(\sigma)}\ \ \lambda_h^{(\sigma)}\right),\ f_i\to\left(f_i^{(\sigma)}\ \ g_i^{(\sigma)}\ \ h_i^{(\sigma)}\right),$$
$$i=x,y,z=1,2,3,\quad \sigma=1,2,3,\ \dots .$$

In the general form, $\sigma$ can be a set of three integer numbers in accordance with the three-dimensional nature of the space. The eigenvectors for an orthonormalized basis in which the tensor kernel can be written in the separable form as

$$S_{ij}(\mathbf{r},\mathbf{r}')=\sum_{\sigma=1}^{\infty}\left[\lambda_f^{(\sigma)}f_i^{(\sigma)}(\mathbf{r})f_j^{(\sigma)}(\mathbf{r}')+\lambda_g^{(\sigma)}g_i^{(\sigma)}(\mathbf{r})g_j^{(\sigma)}(\mathbf{r}')-\left(\lambda_f^{(\sigma)}+\lambda_g^{(\sigma)}\right)h_i^{(\sigma)}(\mathbf{r})h_j^{(\sigma)}(\mathbf{r}')\right],$$
$$\mathbf{h}^{(\sigma)}=\left[\mathbf{f}^{(\sigma)}\mathbf{g}^{(\sigma)}\right],\quad \int\mathbf{f}^{(\sigma)}\mathbf{f}^{(\gamma)}dV=V\delta_{\sigma\gamma},\quad \int\mathbf{g}^{(\sigma)}\mathbf{g}^{(\gamma)}dV=V\delta_{\sigma\gamma},\quad \int\mathbf{h}^{(\sigma)}\mathbf{h}^{(\gamma)}dV=V\delta_{\sigma\gamma},\qquad (20)$$
$$\left(\mathbf{f}^{(\sigma)}\mathbf{g}^{(\gamma)}\right)=0,\qquad \left(\mathbf{f}^{(\sigma)}\mathbf{h}^{(\gamma)}\right)=0,\qquad \left(\mathbf{g}^{(\sigma)}\mathbf{h}^{(\gamma)}\right)=0,$$
$$i,j=1,2,3=x,y,z,\qquad \sigma,\gamma=1,2,3,\ \dots,\ \ \delta_{\sigma\gamma}=\mathrm{diag}(1,\ 1,\ 1,\ \dots).$$

The value of $\lambda_h$ follows from the condition of a traceless kernel, which yields $\lambda_f^{(\sigma)}+\lambda_g^{(\sigma)}+\lambda_h^{(\sigma)}=0$. This form of the kernel offers a possibility of obtaining a series of variational equations with respect to $\sigma$ for determining their eigenvalues and eigenfunctions. For this purpose, the expression for the ITOP (20) should be substituted into the functional of the total free energy density (16) and (17). Expression (20) in the original form satisfies the required properties of tracelessness and symmetry; correspondingly, there is no need to take into account the undetermined Lagrangian multipliers $\eta$ and $\eta_i$ in the variational equations. Instead of them, it is necessary to take into account the property of orthonormalization of the eigenvectors. In the general form, the variation yields

$$\frac{\partial F^{total}}{\partial \lambda_f^{(\sigma)}}=0,\qquad \frac{\partial F^{total}}{\partial \lambda_g^{(\sigma)}}=0,$$

$$\frac{\partial}{\partial x_k}\frac{\partial F}{\partial\left(\partial f_i^{(\sigma)}/\partial x_k\right)}-\frac{\partial F}{\partial f_i^{(\sigma)}}-\sum_\gamma\left(w^{(\sigma\gamma)}f_i^{(\gamma)}+u^{(\sigma\gamma)}g_i^{(\gamma)}\right)=0, \tag{21}$$

$$\frac{\partial}{\partial x_k}\frac{\partial F}{\partial\left(\partial g_i^{(\sigma)}/\partial x_k\right)}-\frac{\partial F}{\partial g_i^{(\sigma)}}-\sum_\gamma\left(v^{(\sigma\gamma)}g_i^{(\gamma)}+u^{(\sigma\gamma)}f_i^{(\gamma)}\right)=0.$$

In the case of separation of the kernel of the form (20), equations for two vectors are sufficient, because the third vector $\mathbf{h}^{(\sigma)}$ is uniquely determined. The first two relations in system (21) are the algebraic equations for determining the eigenvalues $\lambda_f^{(\sigma)}$ and $\lambda_g^{(\sigma)}$; the second pair of equations forms a system of second-order differential equations for the eigenvectors $\mathbf{f}^{(\sigma)}$ and $\mathbf{g}^{(\sigma)}$. The matrix parameters $w^{(\sigma\gamma)}$, $v^{(\sigma\gamma)}$, and $u^{(\sigma\gamma)}$ are the undetermined Lagrangian multipliers for normalization and orthogonality of the vectors from (20). In the case of the variation, these conditions are added to the free energy functional in the form

$$F\to F+\sum_{\sigma,\gamma}\left[\frac{1}{2}\int\left(f_i^{(\sigma)}w^{(\sigma\gamma)}f_i^{(\gamma)}+g_i^{(\sigma)}v^{(\sigma\gamma)}g_i^{(\gamma)}\right)\frac{dV}{V}+f_i^{(\sigma)}u^{(\sigma\gamma)}g_i^{(\sigma)}\right].$$

By virtue of integral normalization, $w^{(\sigma\gamma)}$ and $v^{(\sigma\gamma)}$ are numerical matrices and $u^{(\sigma\gamma)}$ are matrix functions nonlinearly depending on $\mathbf{f}^{(\sigma)}$ and $\mathbf{g}^{(\sigma)}$. To retain the linearity of the differential equations, the orthogonality of the eigenvectors from system (20) can be considered as integral relations satisfied for all values of $\sigma$ and $\gamma$,

$$\int\left(\mathbf{f}^{(\sigma)}\mathbf{g}^{(\gamma)}\right)dV=0,\qquad \int\left(\mathbf{f}^{(\sigma)}\mathbf{h}^{(\gamma)}\right)dV=0,\qquad \int\left(\mathbf{g}^{(\sigma)}\mathbf{h}^{(\gamma)}\right)dV=0.$$

For this definition, the additive to the Lagrangian takes a different form

$$F\to F+\sum_{\sigma,\gamma}\int\left[\frac{1}{2}\left(f_i^{(\sigma)}w^{(\sigma\gamma)}f_i^{(\gamma)}+g_i^{(\sigma)}v^{(\sigma\gamma)}g_i^{(\gamma)}\right)+f_i^{(\sigma)}u^{(\sigma\gamma)}g_i^{(\sigma)}\right]\frac{dV}{V},$$

where $u^{(\sigma\gamma)}$ are now numerical matrices. If the eigenfunctions $\mathbf{f}^{(\sigma)}$ and $\mathbf{g}^{(\sigma)}$ are initially chosen as satisfying the orthogonality condition, $(\mathbf{f}^{(\sigma)}\mathbf{g}^{(\sigma)})\equiv 0$, then the differential equations also become linear because $u^{(\sigma\gamma)}=0$. If harmonics that refer to different $\sigma$ so that they are also orthogonal are initially chosen, $\int \mathbf{f}^{(\sigma)}\mathbf{f}^{(\gamma)}dV\equiv V\delta_{\sigma\gamma}$, $\int \mathbf{g}^{(\sigma)}\mathbf{g}^{(\gamma)}dV\equiv V\delta_{\sigma\gamma}$, then the system of equations is divided into a number of second-order independent differential equations for each harmonic $\sigma$ because the relations $w^{(\sigma\gamma)}=w^{(\sigma)}\delta_{\sigma\gamma}$ and $v^{(\sigma\gamma)}=v^{(\sigma)}\delta_{\sigma\gamma}$ are satisfied in this case.

The scalar order parameters $Q$ and $P$ and the directions of the LC axes $\mathbf{n}$ and $\mathbf{d}$ can be determined as the eigenvalues and eigenvectors from the solutions (if they exist) of the Fredholm equation of the second kind

$$\int S_{ij}(\mathbf{r},\mathbf{r}')n_j(\mathbf{r}')\frac{dV'}{V}=\frac{2}{3}Q(\mathbf{r})n_i(\mathbf{r}),\quad \int S_{ij}(\mathbf{r},\mathbf{r}')d_j(\mathbf{r}')\frac{dV'}{V}=\left[-\frac{1}{3}Q(\mathbf{r})+P(\mathbf{r})\right]d_i(\mathbf{r}). \tag{22}$$

This way of determining the axes and scalar order parameters yields values that differ in the general case from those traditionally introduced (3) on the basis of locality [4, 20, 26, 37-39].

## IV. EIGENVALUES AND EIGENVECTORS IN UNIAXIAL LIQUID CRYSTALS

It is of interest to derive explicit equations for calculating the introduced eigenvalues and eigenvectors. Let us confine ourselves to considering a uniaxial LC without any external effects. The integral tensor of the order parameter is written with due allowance for the tracelessness in the form

$$S_{ij}(\mathbf{r},\mathbf{r}')=\sum_\sigma\lambda^{(\sigma)}\left[f_i^{(\sigma)}(\mathbf{r})f_j^{(\sigma)}(\mathbf{r}')-\frac{\delta_{ij}}{3}f_k^{(\sigma)}(\mathbf{r})f_k^{(\sigma)}(\mathbf{r}')\right],\qquad \int\mathbf{f}^{(\sigma)}\mathbf{f}^{(\gamma)}\frac{dV}{V}=\delta_{\sigma\gamma}. \tag{23}$$

Expression (23) should be substituted into Eq. (17); In this case, it is sufficient to consider terms with differentiation only with respect to the coordinates **r**, because taking into account terms with differentiation with respect to **r**' yields the same result by virtue of symmetry. After the above-mentioned substitution, integration with respect to **r**' becomes possible, and the final result can be written in the quadratic form:

$$F_{el}(\mathbf{r}) = \sum_{\sigma} \left(\lambda^{(\sigma)}\right)^2 \varphi^{(\sigma)}(\mathbf{r}),$$

$$\varphi^{(\sigma)}(\mathbf{r}) = \frac{1}{2}\left[\begin{array}{c} \frac{\partial f_k^{(\sigma)}}{\partial x_i} A_{ij}^{(\sigma)} \frac{\partial f_k^{(\sigma)}}{\partial x_j} + \frac{\partial f_i^{(\sigma)}}{\partial x_k} B_{ij}^{(\sigma)} \frac{\partial f_j^{(\sigma)}}{\partial x_k} + \frac{\partial f_k^{(\sigma)}}{\partial x_k} \bar{C}_{ij}^{(\sigma)} \frac{\partial f_i^{(\sigma)}}{\partial x_j} + \frac{\partial f_k^{(\sigma)}}{\partial x_j} \bar{D}_{ij}^{(\sigma)} \frac{\partial f_i^{(\sigma)}}{\partial x_k} + \\ + q\left(2\varepsilon_{ijk} H_{kl}^{(\sigma)} f_i^{(\sigma)} \frac{\partial f_l^{(\sigma)}}{\partial x_j} + q H_{ij}^{(\sigma)} f_j^{(\sigma)} f_i^{(\sigma)}\right) \end{array}\right]. \tag{24}$$

To simplify the form of this expression, only two terms with the coefficients $M_{1q} = M_{5q} \neq 0$ and $M_{2q} = M_{6q} \neq 0$ are left in the last term. If necessary, other terms can be taken into account by using the permutation of the subscripts $l \leftrightarrow i$. Integration gives rise to numerical matrix coefficients, which are defined as

$$\begin{gathered} a_{ij}^{(\sigma)} = \int f_i^{(\sigma)} f_j^{(\sigma)} \frac{dV}{V}, \\ A_{ij}^{(\sigma)} = \left(M_2 + M_3\right) a_{ij}^{(\sigma)},\ B_{ij}^{(\sigma)} = M_1 \delta_{ij} + M_4 a_{ij}^{(\sigma)},\ \bar{C}_{ij}^{(\sigma)} = M_7 \delta_{ij} + M_5 a_{ij}^{(\sigma)}, \\ \bar{D}_{ij}^{(\sigma)} = M_8 \delta_{ij} + M_6 a_{ij}^{(\sigma)},\ C_{ij}^{(\sigma)} = \frac{1}{2}\left(\bar{C}_{ij}^{(\sigma)} + \bar{D}_{ij}^{(\sigma)}\right) = \frac{1}{2}\left[\left(M_7 + M_8\right)\delta_{ij} + \left(M_5 + M_6\right) a_{ij}^{(\sigma)}\right], \\ H_{ij}^{(\sigma)} = M_{1q} \delta_{ij} + M_{2q} a_{ij}^{(\sigma)}. \end{gathered} \tag{25}$$

In system (25), we additionally introduce $C_{ij}^{(\sigma)} = \left(\bar{C}_{ij}^{(\sigma)} + \bar{D}_{ij}^{(\sigma)}\right)/2$, which will be required later in writing the variational equations. The Landau-de Gennes potential (17) takes the form

$$\begin{gathered} U = \sum_{\sigma} \left(\lambda^{(\sigma)}\right)^2 f_i^{(\sigma)} f_j^{(\sigma)} \left[\begin{array}{c} \frac{1}{2}\left(\alpha_1 \delta_{ij} + \alpha_2 a_{ij}^{(\sigma)}\right) - \lambda^{(\sigma)}\left(\beta_1 \delta_{ij} a_{kl}^2 + \beta_2 a_{ik} a_{kj}\right) + \\ + \left(\lambda^{(\sigma)}\right)^2 \left(\gamma_1 \delta_{ij} a_{kl}^3 + \gamma_2 a_{il} a_{lk} a_{kj}\right) \end{array}\right], \\ U^{total} = \sum_{\sigma} \left(\lambda^{(\sigma)}\right)^2 \left[\frac{\alpha^{(\sigma)}}{2} - \beta^{(\sigma)} \lambda^{(\sigma)} + \gamma^{(\sigma)} \left(\lambda^{(\sigma)}\right)^2\right], \end{gathered} \tag{26}$$

where the coefficients $\alpha$, $\beta$, and $\gamma$ are calculated via the matrix elements $a_{ij}$ as

$$\begin{gathered} \alpha^{(\sigma)} = \alpha_1 + \alpha_2 \left(a_{ij}^{(\sigma)}\right)^2, \quad \beta^{(\sigma)} = \beta_1 \left(a_{kl}^{(\sigma)}\right)^2 + \beta_2 a_{ik}^{(\sigma)} a_{kj}^{(\sigma)} a_{ij}^{(\sigma)}, \\ \gamma^{(\sigma)} = \gamma_1 \left(a_{kl}^{(\sigma)}\right)^3 + \gamma_2 a_{il}^{(\sigma)} a_{kl}^{(\sigma)} a_{kj}^{(\sigma)} a_{ij}^{(\sigma)}. \end{gathered} \tag{27}$$

In Eq. (26), the eigenvalues of the ITOP operator are taken as invariants (18). Formally, the last term $q^2 H_{ij}^{(\sigma)} f_j^{(\sigma)} f_i^{(\sigma)}$ for the cholesteric additive in Eq. (24) can be transferred to potential (26) because this term is independent of the derivatives.

The variational equations (21) for the eigenfunctions $\mathbf{f}^{(\sigma)}$ obtained from relations (24), (25), and (26) take the form

$$\frac{\partial}{\partial x_k}\left[A_{kj}^{(\sigma)} \frac{\partial f_i^{(\sigma)}}{\partial x_j} + B_{ij}^{(\sigma)} \frac{\partial f_j^{(\sigma)}}{\partial x_k} + C_{ij}^{(\sigma)} \frac{\partial f_k^{(\sigma)}}{\partial x_j} + C_{jk}^{(\sigma)} \frac{\partial f_j^{(\sigma)}}{\partial x_i} + \left(\varepsilon_{lkj} H_{ji}^{(\sigma)} - \varepsilon_{ikj} H_{jl}^{(\sigma)}\right) f_l^{(\sigma)}\right] = \sum_{\gamma} w^{(\sigma\gamma)} f_i^{(\gamma)}, \tag{28}$$

where the matrix coefficients $w^{(\sigma\gamma)}$ are determined from the normalization condition

$$\int \mathbf{f}^{(\sigma)}\mathbf{f}^{(\gamma)}\frac{dV}{V}=\delta_{\sigma\gamma}.$$

The undetermined Lagrangian multipliers $w^{(\sigma\gamma)}$ include linear (with respect to $\mathbf{f}^{(\sigma)}$) terms arising in the case of the variation of the Landau-de Gennes potential (26).

For obtaining equations determining the eigenvalues from Eq. (21), it is necessary to equate the derivative of the total free energy $F^{total}$ (16) with respect to $\lambda^{(\sigma)}$ to zero. The result has the form of the third-order algebraic equation

$$\lambda^{(\sigma)}\left[4\gamma^{(\sigma)}\left(\lambda^{(\sigma)}\right)^2-3\beta^{(\sigma)}\lambda^{(\sigma)}+2\Phi^{(\sigma)}+\alpha^{(\sigma)}\right]=0. \qquad (29)$$

Three roots of the cubic equation have the form

$$\lambda_0^{(\sigma)}=0,\qquad \lambda_{1,2}^{(\sigma)}=\frac{3\beta^{(\sigma)}\pm\sqrt{9\left(\beta^{(\sigma)}\right)^2-16\gamma^{(\sigma)}\left(2\Phi^{(\sigma)}+\alpha^{(\sigma)}\right)}}{8\gamma^{(\sigma)}},$$

where a variable $\Phi^{(\sigma)}$ is introduced, which can be identified with the energy of the harmonic $\sigma$ (24):

$$\Phi^{(\sigma)}=\int \varphi^{(\sigma)}dV.$$

The resultant roots characterize the phase states of each harmonic; finally, after calculation of the free energy (16), (24), and (26), these roots describe the phase state of the deformed LC. As usually, the zero root of Eq. (29) corresponds to the isotropic state of the NLC. Two other roots describe the phase transitions. It should be noted that the form of the cubic equation (29) is equivalent to the equation for the scalar order parameter (3), in the case of its local determination [1–3, 6]. Various issues associated with the aggregate states of liquid crystals are not considered in detail in this work. Using the analogy with the ideology of the classical theory of phase transitions in liquid crystals, it can be expected that the coefficient $\alpha^{(\sigma)}$ near the phase transition should be linearly dependent on the temperature $T$ as

$$\alpha^{(\sigma)}\sim T-T_c^{(\sigma)},$$

where $T_c^{(\sigma)}$ is the temperature of the phase transition for each harmonic $\sigma$.

Thus, we can state that a generalized ITOP-based phenomenological approach for the LC description is proposed, and a method of its implementation in the form of solving an infinite series of linear differential equations and nonlinear algebraic equations for harmonics that are eigenvectors and eigenvalues of the introduced integral operator of the tensor order parameter.

## V. LOCAL APPROXIMATION. COMPARISON WITH THE OSEEN-FRANK THEORY

If the director is immediately introduced into consideration, then the order parameter for uniaxial LCs can be redefined as

$$S_{ij}(\mathbf{r},\mathbf{r}')=S(\mathbf{r},\mathbf{r}')\left[n_i(\mathbf{r})n_j(\mathbf{r}')-\frac{\delta_{ij}}{3}n_k(\mathbf{r})n_k(\mathbf{r}')\right],$$
$$S_{ij}(\mathbf{r},\mathbf{r}')=S_{ji}(\mathbf{r}',\mathbf{r})\ \rightarrow\ S(\mathbf{r},\mathbf{r}')=S(\mathbf{r}',\mathbf{r}). \qquad (30)$$

It follows from definition (30) that the director $\mathbf{n}(\mathbf{r})$ is a unit eigenvector of the operator $S_{ij}$ ($\mathbf{r}$,$\mathbf{r}'$) if the scalar function is sufficiently narrow or close to the delta function

$$S(\mathbf{r},\mathbf{r}')\approx S(\mathbf{r})\delta(\mathbf{r}-\mathbf{r}'), \qquad (31)$$

then we have

$$\int S_{ij}(\mathbf{r},\mathbf{r}')n_j(\mathbf{r}')\frac{dV'}{V}\xrightarrow[S(\mathbf{r},\mathbf{r}')\to S(\mathbf{r})\delta(\mathbf{r}-\mathbf{r}')]{}\frac{2}{3}Q(\mathbf{r})n_i(\mathbf{r}),\quad Q(\mathbf{r})=S(\mathbf{r},\mathbf{r})\equiv S(\mathbf{r}),$$

where $2Q/3$ is the approximate eigenvalue if condition (31) is satisfied. The physical meaning of the condition at which the scalar two-point order parameter tends to the delta function is drastic reduction of the influence of the neighboring (at the point $\mathbf{r}'$) domains of the point $\mathbf{r}$ on the LC state at the point **r**. Moreover, this condition should not be violated in the case of deformations or other actions on the LC whose size should be much greater than the width of the kernel $S(\mathbf{r},\mathbf{r}')$. Based on these considerations, it may be assumed that the scalar kernel in the general case should satisfy the inequality $-1/2 \le S(\mathbf{r},\mathbf{r}') \le 1$, like $-1/2 \le Q(\mathbf{r}) \le 1$ for local determination.

For comparisons with the classical Oseen-Frank theory on the basis of Eq. (30) under condition (31) it is necessary to calculate the free energy density (17) with ignored derivatives with respect to the primed coordinate **r**'. The corresponding calculations can be formally written in the form of integrals with the $\delta$ –function:

$$F_{el}(\mathbf{r}) = \frac{1}{2}\int \begin{bmatrix} M_1 S_{ij,k} S_{ij,k} + M_2 S_{ik,k} S_{ij,j} + M_3 S_{ij,k} S_{ik,j} + M_4 S_{ji,k} S_{ij,k} + M_5 S_{ki,k} S_{ij,j} + \\ + M_6 S_{ji,k} S_{ik,j} + M_7 S_{ki,k} S_{ji,j} + M_8 S_{ji,k} S_{ki,j} + M_q q\left(2\varepsilon_{ijk} S_{il} S_{kl,j} + q S_{ij} S_{ji}\right) \end{bmatrix} \delta(\mathbf{r}-\mathbf{r}')\, dV',$$

$$U(\mathbf{r}) = \int \left( \frac{\alpha}{2} S_{ij} S_{ji} - \beta S_{ij} S_{jk} S_{ki} + \gamma S_{ij} S_{ji} S_{kl} S_{lk} \right) \delta(\mathbf{r}-\mathbf{r}')\, dV', \tag{32}$$

$$S_{ij,k} = \frac{\partial Q}{\partial x_k}\left( n_i n_j' - \frac{\delta_{ij}}{3} n_l n_l' \right) + Q\left( n_j' \frac{\partial n_i}{\partial x_k} - \frac{\delta_{ij}}{3} n_l' \frac{\partial n_l}{\partial x_k} \right), \quad Q = Q(\mathbf{r}), \quad n_i = n_i(\mathbf{r}), \quad n_i' = n_i(\mathbf{r}').$$

In writing Eqs. (32), we omitted terms with derivatives of the $\delta$-function in the order parameter (31) because the allowance for these derivatives would lead to the emergence of higher-order terms, i.e., second derivatives and first derivatives with powers greater than the second one. Such terms can be neglected by virtue of the adopted approximation of a narrow kernel and large deformations. Combinations that are independent in the case of approximation (31) are left in the Landau-de Gennes potential and in the cholesteric term. The sequence of procedures is very important in calculations: first the derivatives of the order parameter (30) are calculated and then integration is performed. If these procedures are performed in the opposite order, the result is identical to that obtained in [4, 20, 38, 39] and in [26, 37] for $L_4$, $L_5$, … = 0, namely $K_1 = K_3$. This violation of commutativity occurs owing to symmetry with respect to the subscripts due to integration performed at the first stage, $S_{ij}(\mathbf{r},\mathbf{r}) = S_{ji}(\mathbf{r},\mathbf{r})$, and the final result is qualitatively different. Calculations (see more details in Appendix) yield the following expressions similar to Eqs. (8) and (9):

$$F = F_0 + F_{el}(\mathbf{r}) + U(\mathbf{r}), \quad F_{el}(\mathbf{r}) = \frac{1}{2}\left( \frac{\partial Q}{\partial x_i} D_{ij} \frac{\partial Q}{\partial x_j} + P_k \frac{\partial Q^2}{\partial x_k} \right) + F_{el}^{Frank}. \tag{33}$$

The Frank constants of elasticity $K_i = k_i Q^2,\ i = 1,\ldots,4$ and the parameters $D_\perp$, $D_\parallel$, $\mu_1$, $\mu_3$, $A$, $B$, and $C$ are uniquely expressed via the coefficients $M_j,\ j = 1,\ldots,8,$ $M_q = 3M_l/2$, $\alpha$, $\beta$, and $\gamma$ by the relations

$$D_\parallel = \frac{2}{3}(M_1 + M_4) + \frac{4}{9}(M_2 + M_3 + M_5 + M_6 + M_7 + M_8),$$

$$D_\perp = \frac{2}{3}(M_1 + M_4) + \frac{1}{9}(M_2 + M_3 + M_5 + M_6 + M_7 + M_8),$$

$$D_a = D_\parallel - D_\perp = \frac{1}{3}(M_2 + M_3 + M_5 + M_6 + M_7 + M_8), \tag{34}$$

$$\mu_1 = \frac{1}{3}\left( -M_3 + M_5 - \frac{1}{2}M_6 + 2M_7 \right), \quad \mu_3 = \frac{1}{3}\left( -M_2 - \frac{1}{2}M_5 + M_6 + 2M_8 \right),$$

$$\mu_1 + \mu_3 = \frac{1}{3}\left[ -M_2 - M_3 + \frac{1}{2}(M_5 + M_6) + 2(M_7 + M_8) \right],$$

$$k_1 \equiv \frac{K_1}{Q^2} = M_1 + M_7 + M_8, \quad k_2 \equiv \frac{K_2}{Q^2} = M_1, \quad k_3 \equiv \frac{K_3}{Q^2} = M_1 + M_2 + M_3, \quad \frac{K_2 + K_4}{Q^2} = M_8,$$

$$A = \frac{2}{3}\alpha, \;\; B = \frac{2}{9}\beta, \;\; C = \frac{4}{9}\gamma.$$

In contrast to the local approach [4, 20, 26, 37-39], the resultant general expressions ensure the validity of the inequalities $K_1 \neq K_2 \neq K_3 \neq K_4$ in the quadratic form (for the derivatives of the order parameter) of the elastic part of the free energy density (32). This is a consequence of the absence of symmetry only by the subscripts, $S_{ij}(\mathbf{r},\mathbf{r}') \neq S_{ji}(\mathbf{r},\mathbf{r}')$. The constants obtained in Eq. (34) (except the coefficients of the Landau-de Gennes potential) can be conditionally divided into bulk and surface ones. Three quantities, $K_2+K_4$ and flex coefficients $\mu_1$ and $\mu_3$ can be considered as surface ones. Respectively, five constants $D_{\|}$, $D_{\perp}$ and $K_1$, $K_2$, $K_3$ will be "purely" bulk.

Formulation (30) and approximation (31) in the general form lead to the emergence of eight constants; four constants of them are new; they have to be determined from additional experimental investigations, e.g., similar to those performed in [15-17] for thermo-orientation effects. A preliminary analysis of these publications allows us to conclude that there may be some proportionality relations. The tensor is $D_{ij} \sim \kappa_{ij}$, where $\kappa_{ij}$ is the thermal conductivity tensor [15, 16]. Based on the form of the vector $\mathbf{P}$ (33), which is similar to the flexoelectric vector, and on the results of [17], it may be assumed that $\mu_1 \sim e_1$ and $\mu_3 \sim e_3$.

The quadratic dependence of the elasticity coefficients on the scalar order parameter ($K_i = k_i Q^2$, $\partial k_i/\partial Q = 0$) offers a possibility of writing the variational equations for the order parameters and for the director in a comparatively simple form (in contrast to Eqs. (8)–(11), where the variation with respect to $Q$ yields more complicated relations because of Eq. (13)):

$$\nabla \cdot \left[ D_{\perp} \nabla Q + D_a \mathbf{n} \left( \mathbf{n} \cdot \nabla Q \right) \right] + Q \left( \nabla \cdot \mathbf{P} - 2\Lambda \right) = U_Q,$$

$$\begin{gathered} \nabla \left( k_1 Q^2 \nabla \cdot \mathbf{n} + \mu_1 Q \mathbf{n} \cdot \nabla Q \right) - \\ - \nabla \times \left\{ Q^2 \left[ \left( k_2 - k_3 \right) \mathbf{n} \left( \mathbf{n} \cdot \nabla \times \mathbf{n} \right) + k_3 \nabla \times \mathbf{n} \right] + \mu_3 Q \mathbf{n} \times \nabla Q + k_2 Q^2 q \mathbf{n} \right\} - \\ - Q^2 \left( k_2 - k_3 \right) \left( \mathbf{n} \cdot \nabla \times \mathbf{n} \right) \nabla \times \mathbf{n} - D_a \nabla Q \left( \mathbf{n} \cdot \nabla Q \right) - \mu_1 Q \nabla Q \nabla \cdot \mathbf{n} + \mu_3 Q \nabla Q \times \nabla \times \mathbf{n} - w \mathbf{n} = 0, \end{gathered} \tag{35}$$

where

$$\Lambda = F_{el}^{Frank} / Q^2, \qquad \nabla \cdot \mathbf{P} = \left( \mu_1 + \mu_3 \right) \left[ \left( \nabla \cdot \mathbf{n} \right)^2 + \mathbf{n} \cdot \nabla \left( \nabla \cdot \mathbf{n} \right) \right],$$

$U_Q = \partial U/\partial Q$, and $w$ is the undetermined Lagrangian multiplier, which ensures the unit character of the director. At $Q = const$ and $q = 0$, the equilibrium equation coincides with the known general equation [3] (all elastic constants are different, $K_1 \neq K_2 \neq K_3$) for determining the director $\mathbf{n}$ in NLCs. As was already noted in Section 2, Eq. (35), similar to Eqs. (8) and (9), involves the effect of "induced biaxiality" at $\mathbf{n} \times \mathbf{P} \neq 0$, $\mathbf{n} \times \nabla Q \neq 0$. From the form of the equation for the director, there follows one more important consequence: the gradient of the order parameter can affect the LC director orientation like an electric field. This may be realized, e.g., in the presence of an external temperature gradient $\nabla Q = (\partial Q/\partial T)\nabla T$, which was observed in NLCs [15-17], and also in the form of "self-action" in the smectic order where $\nabla Q = (\partial Q/\partial \rho)\nabla \rho$ [1-3].

Emergence of the second preferred direction, as was already mentioned in Section 2, formally is not completely correct, because it contradicts the original assumption that the LC medium is uniaxial. In particular, if collinearity of the vectors $\mathbf{P}$ and $\mathbf{n}$, $\mathbf{P} \times \mathbf{n} = 0$ is required additionally in Eq. (33) then the coefficient $\mu_3$ is to vanish, or

$$\mathbf{P} \times \mathbf{n} = 0 \rightarrow \begin{cases} \mathbf{P}_{\|} = \mu_1 \mathbf{n} \nabla \cdot \mathbf{n} \neq 0, \\ \mathbf{P}_{\perp} = -\mu_3 \mathbf{n} \times \nabla \times \mathbf{n} = 0, \end{cases} \rightarrow \mu_3 = 0.$$

Such a condition leads to the equality

$$\mu_1 = \mu_1 + \mu_3 = \frac{1}{3}\left[-M_2 - M_3 + \frac{1}{2}\left(M_5 + M_6\right) + 2\left(M_7 + M_8\right)\right],$$

instead of obtaining the flex coefficients from Eq. (34). If it is taken into account then there are only five linearly independent coefficients so that it can be assumed that $M_3 = M_6 = M_7 = 0$. Naturally, it is not necessary if we originally take biaxiality in Eq. (3) at $Q \neq 0$ and $P \neq 0$ into account. The equation obtained directly at variation of Eq. (19) can serve as a confirmation. For NLC it can written as

$$\frac{\partial}{\partial x_k}\left\{\frac{\partial}{\partial x_k}\left(M_1 S_{ij} + M_4 S_{ji}\right) + \left(M_2 + M_3\right)\frac{\partial S_{ik}}{\partial x_j} + \frac{M_5 + M_6}{2}\left(\frac{\partial S_{ki}}{\partial x_j} + \frac{\partial S_{jk}}{\partial x_i}\right) + \left(M_7 + M_8\right)\frac{\partial S_{kj}}{\partial x_i}\right\} +$$

$$+\frac{\partial}{\partial x'_k}\left\{\frac{\partial}{\partial x'_k}\left(M_1 S_{ij} + M_4 S_{ji}\right) + \left(M_2 + M_3\right)\frac{\partial S_{ik}}{\partial x'_j} + \frac{M_5 + M_6}{2}\left(\frac{\partial S_{ki}}{\partial x'_j} + \frac{\partial S_{jk}}{\partial x_i}\right) + \left(M_7 + M_8\right)\frac{\partial S_{kj}}{\partial x'_i}\right\} =$$

$$= \frac{\partial U}{\partial S_{ij}} - \eta\delta_{ij} - \eta_{ij}.$$

The above-derived nonlinear differential equations (35) for the director and scalar order parameter can be adopted as a basis for the description of a uniaxial LC in the locality or short-range approximation (31). This approach is also extended to biaxial LCs. In the case of a sufficiently compact kernel, it is close (identical in the locality limit) to the traditional description of LCs with the use of the director or the direction of the axes of the local tensor order parameter. The advantage of the method is the possibility of solving the problem of inequality of the Frank constants within the framework of the quadratic equation of elastic energy (33), which simplifies the equilibrium equations of the LC structure.

## VI. EXAMPLES

### A. One-dimensional S-, T-, and B-deformations

It is of interest to consider some particular cases on the basis of the ITOP concept and to compare the result with that obtained on the basis of the local definition of the order parameter. First of all, such a comparison is needed to derive relations between the constants $M_{1\text{-}8}$ introduced here with the Frank constants $K_{1,2,3}$. The simplest option is one-dimensional variants with the geometry determining S-, T-, and B-deformations in NLCs [1–3] with stiff boundary conditions. Let us assume that an NLC ($q = 0$) occupies a certain space shaped as an infinite layer in the $x$ and $y$ directions with a finite thickness $L$ in the $z$ direction. The functional $\varphi^{(\sigma)}$ for the eigenvectors of the ITOP from Eq. (24) takes the form

$$\varphi^{(\sigma)}(z) = \frac{1}{2}\left[\frac{\partial f_i^{(\sigma)}}{\partial z}\left(A_{zz}^{(\sigma)}\delta_{ij} + B_{ij}^{(\sigma)}\right)\frac{\partial f_j^{(\sigma)}}{\partial z} + 2\frac{\partial f_i^{(\sigma)}}{\partial z}C_{iz}\frac{\partial f_z^{(\sigma)}}{\partial z}\right], \tag{36}$$

The variational equations obtained on the basis of Eq. (36) are written in the form of the following system:

$$\begin{cases} \dfrac{\partial}{\partial z}\left[\left(A_{zz}^{(\sigma)}\delta_{ij} + B_{ij}^{(\sigma)}\right)\dfrac{\partial f_j^{(\sigma)}}{\partial z} + C_{iz}\dfrac{\partial f_z^{(\sigma)}}{\partial z}\right] = \sum_\gamma w^{(\sigma\gamma)} f_i^{(\gamma)}, & i = x, y, \\ \dfrac{\partial}{\partial z}\left[\left(B_{zj}^{(\sigma)} + C_{zj}^{(\sigma)}\right)\dfrac{\partial f_j^{(\sigma)}}{\partial z} + \left(A_{zz}^{(\sigma)} + C_{zz}^{(\sigma)}\right)\dfrac{\partial f_z^{(\sigma)}}{\partial z}\right] = \sum_\gamma w^{(\sigma\gamma)} f_z^{(\gamma)}, & i = z. \end{cases} \tag{37}$$

For B-deformation, we assume that

$$\mathbf{f}^{(\sigma)} = \left(0,\ f_y^{(\sigma)},\ f_z^{(\sigma)}\right),$$

with the boundary conditions

$$\left(f_y^{(\sigma)},\ \ f_z^{(\sigma)}\right)_{z=0,L} = (0,\ \ 1).$$

For comparisons with the results of the classical Frank theory, we consider small deviations and seek for the solution in the form of one harmonic

$$\mathbf{f}^{(1)} = \left(f_y,\ \ f_z\right) \approx \left(f_y,\ \ 1\right),\quad f_y << 1,\quad f_z \approx 1.$$

In this approximation, the Lagrangian multipliers are equal to zero, and the matrix elements (25) have the form

$$A_{ij}^{(1)} = \left(M_2 + M_3\right) \int \begin{pmatrix} f_y^2 & f_y \\ f_y & 1 \end{pmatrix} \frac{dz}{L} \underset{f_y \to 0}{\to} \left(M_2 + M_3\right) \begin{pmatrix} 0 & 0 \\ 0 & 1 \end{pmatrix},$$

$$B_{ij}^{(1)} \underset{f_y \to 0}{\to} \begin{pmatrix} M_1 & 0 \\ 0 & M_1 + M_4 \end{pmatrix},\qquad C_{ij}^{(1)} \underset{f_y \to 0}{\to} \frac{1}{2} \begin{pmatrix} M_7 + M_8 & 0 \\ 0 & M_7 + M_8 + M_5 + M_6 \end{pmatrix}.$$

Substituting the resultant expressions into Eqs. (36) and (37), we find

$$F_{el}(\mathbf{r}) = \sum_\sigma \left(\lambda^{(\sigma)}\right)^2 \varphi^{(\sigma)}(\mathbf{r}) \approx \left(\lambda^{(1)}\right)^2 \varphi^{(1)}(\mathbf{r}) = \frac{M_1 + M_2 + M_3}{2} \left(\lambda^{(1)}\right)^2 \left(\frac{\partial f_y}{\partial z}\right)^2 \overset{Var}{\to}$$

$$\to \quad \left(M_1 + M_2 + M_3\right) \frac{\partial^2 f_y}{\partial z^2} = 0.$$

It follows from these expressions that the following relation is valid for B-deformation:

$$K_3 \approx \left(\lambda^{(1)}\right)^2 \left(M_1 + M_2 + M_3\right).$$

Performing similar considerations for other types of deformation and taking into account Eq. (29), we finally obtain

$$K_1 \approx \left(\lambda^{(1)}\right)^2 \left(M_1 + M_7 + M_8\right),\quad K_2 \approx \left(\lambda^{(1)}\right)^2 M_1,\quad K_3 \approx \left(\lambda^{(1)}\right)^2 \left(M_1 + M_2 + M_3\right),$$

$$\lambda^{(1)} \approx \frac{3\beta^{(1)} + \sqrt{9\left(\beta^{(1)}\right)^2 - 16\gamma^{(1)}\alpha^{(1)}}}{8\gamma^{(1)}}. \tag{38}$$

Equation (38) agrees well with similar relations (34) derived in the approximation of a short-range action or a narrow kernel.

### B. One-dimensional LC, local approximation

Let us consider a one-dimensional case, where the director and order parameter depend only on the *z* coordinate:

$$\mathbf{n} = (\sin\theta\cos\varphi,\ \sin\theta\sin\varphi,\ \cos\theta),\ \theta = \theta(z),\ \varphi = \varphi(z),\ Q(z),\ -\frac{L}{2} \le z \le \frac{L}{2}. \tag{39}$$

The free energy density functional (33), (34) without the isotropic term with allowance for Eq. (39) is written as

$$F = \frac{1}{2}\left\{DQ_z^2 - \mu\theta_z QQ_z \sin 2\theta + Q^2\left[f_{13}\theta_z^2 + \sin^2\theta\varphi_z\left(f_{23}\varphi_z - 2k_2 q\right)\right] + k_2 q^2\right\} + U(Q),$$

$$f_{13}(\theta) = k_1 \sin^2\theta + k_3 \cos^2\theta,\qquad f_{23}(\theta) = k_2 \sin^2\theta + k_3\cos^2\theta, \tag{40}$$

$$\mu = \mu_1 + \mu_3,\qquad D = D_\parallel \sin^2\theta + D_\perp \cos^2\theta.$$

The complex $k_2 q^2 Q^2/2$ can be formally introduced into the Landau-de Gennes potential; the subscript *z* means the derivative with respect to *z*. The variational equations from system (40) for the functions $Q$, $\theta$, and $\varphi$ have the form

$$\begin{cases} \dfrac{d}{dz}\left(DQ_z\right)-Q\left\{\mu\dfrac{\sin 2\theta}{2}\theta_{zz}+g_{13}\theta_z^2+S_\theta^2\varphi_z\left(f_{23}\varphi_z-2k_2q\right)\right\}=k_2q^2Q+U_Q, \\ \dfrac{d}{dz}\left(Q^2f_{13}\theta_z\right)-\dfrac{\sin 2\theta}{2}\left\{\mu QQ_{zz}+\left(D_a+\mu\right)Q_z^2+\left[k_{13}\theta_z^2+\left(f_{23}+k_{23}\sin^2\theta\right)\varphi_z^2-2k_2q\varphi_z\right]Q^2\right\}=0, \\ \dfrac{d}{dz}\left[Q^2\sin^2\theta\left(f_{23}\varphi_z-k_2q\right)\right]=0, \end{cases} \tag{41}$$

$$g_{13}=\left(k_1-\mu\right)S_\theta^2+\left(k_3+\mu\right)C_\theta^2.$$

where the subscript $Q$ means the derivative with respect to $Q$. The last equation for the angle $\varphi$ of system (41) can be integrated once, which yields

$$\frac{d\varphi}{dz}=\frac{k_2}{f_{23}}\left(\frac{M}{Q^2\sin^2\theta}+q\right),$$

where $M$ is the constant of integration. Substitution of the derivative of the azimuthal angle $\varphi$ into the free energy functional yields the expression

$$F=D\frac{Q_z^2}{2}-\mu\frac{\sin 2\theta}{2}\theta_zQQ_z+\left[f_{13}\theta_z^2+k_2q^2\left(1-\sin^2\theta\frac{k_2}{f_{23}}\right)\right]\frac{Q^2}{2}+\frac{1}{2}\frac{k_2^2}{f_{23}}\frac{M^2}{Q^2\sin^2\theta}+U\left(Q\right),$$

From Eqs. (40) and the first integral for the derivative of the azimuthal angle $\varphi$, we can write one more conservation integral in the form of the momentum (zz are the components of the tension tensor [41]) of the system

$$H=const=\frac{\partial F}{\partial Q_z}Q_z+\frac{\partial F}{\partial\theta_z}\theta_z+\frac{\partial F}{\partial\varphi_z}\varphi_z-F=$$

$$=\frac{1}{2}\left[DQ_z^2-\mu\theta_zQQ_z\sin 2\theta+\left(f_{13}\theta_z^2+f_{23}\varphi_z^2\sin^2\theta-k_2q^2\right)Q^2\right]-U\left(Q\right)=$$

$$=\frac{1}{2}\left\{DQ_z^2-\mu\theta_zQQ_z\sin 2\theta+\left[f_{13}\theta_z^2+\frac{k_2^2}{f_{23}}\left(\frac{M}{Q^2\sin\theta}+q\sin\theta\right)^2-k_2q^2\right]Q^2\right\}-U\left(Q\right).$$

In the general form, it is reasonable to solve system (41) by numerical methods, despite its one-dimensionality and steadiness.

For a cholesteric LC (CLC), are simplified after the substitution $\theta=\pi/2$ and take the form

$$D_{\|}\frac{dQ_z}{dz}-k_2\varphi_z\left(\varphi_z-2q\right)Q=V_Q,\qquad \frac{d}{dz}\left[Q^2\left(\varphi_z-q\right)\right]=0,$$
$$Q\big|_{z=\pm L/2}=1,\qquad \varphi\big|_{z==\pm L/2}=0. \tag{42}$$

In the case of boundary conditions correlated with the helix pitch $q=2\pi/\lambda$ ($\lambda$ is the helix pitch in the CLC and $n\lambda/2=L$, where $n$ is an integer number) for the azimuthal angle $\varphi$, the second equation yields a simple particular solution $\varphi=qz+\varphi_0$. The first equation can be integrated, which yields the implicit relation

$$\int\frac{dQ}{\sqrt{U\left(Q\right)+U_0}}=\pm\left(\sqrt{\frac{2}{D_{\|}}}z+C\right),$$

where $\varphi_0$, $C$, and $U_{L/2}$ are the constant of integration, and $U_{L/2}$ is the value of the potential at the center of the layer. If we retain only the quadratic term $U=AQ^2/2$ in the Landau-de Gennes potential, then the equation for the scalar order parameter becomes linear and can be integrated explicitly:

$$Q=Q_{L/2}\mathrm{ch}\left(\beta z\right),\quad \beta=\sqrt{\frac{|A|}{D_{\|}}},\quad Q_{L/2}=\mathrm{ch}^{-1}\left(\sqrt{\frac{|A|}{D_{\|}}}\frac{L}{2}\right). \tag{43}$$

It is of interest that Eqs. (42) directly yield the "helical" solution ($\varphi = qz$), whereas the equilibrium equations for a one-dimensional CLC in the Frank-de Gennes theory yield $\partial^2\varphi/\partial z^2 = 0 \rightarrow \varphi = const \cdot z + \varphi_0$, and an additional analysis is needed to obtain a physically correct solution in the form of a helical structure with a constant pitch ($const = q$) [1-3].

For an NLC with, e.g., B-deformation, the equilibrium equations (41) in the linear approximation ($\theta << 1$, $\varphi = 0$) take the form

$$D_\perp \frac{d^2Q}{dz^2} = U_Q, \qquad k_3 \frac{d}{dz}\left(Q^2 \frac{d\theta}{dz}\right) - \theta\left[\mu Q \frac{d^2Q}{dz^2} + (\mu + D_a)\left(\frac{dQ}{dz}\right)^2\right] = 0,$$

$$Q\big|_{z=\pm L/2} = 1, \qquad \theta\big|_{z=\pm L/2} = 0.$$

If, for certainty, we use the solution of the first equation for the quadratic approximation in the Landau-de Gennes potential, similar to Eq. (43) with the replacement $D_{\|} \rightarrow D_\perp$, we obtain one equation for the polar angle $\theta$:

$$k_3 \frac{d}{dx}\left(\mathrm{ch}^2(\beta x)\frac{d\theta}{dz}\right) - \beta^2\theta\left[\mu\,\mathrm{ch}^2(\beta x) + (\mu + D_a)\,\mathrm{sh}^2(\beta x)\right] = 0,$$

$$\theta\big|_{x=\pm L/2} = 0, \qquad \beta = \sqrt{\frac{|A|}{D_\perp}}. \tag{44}$$

This equation differs from the result of the classical Frank theory, where $Q = const$ and only one term of the form $K_3\partial^2\theta/\partial z^2 = 0$ is left. In addition to the spatial dependence of the elasticity coefficient $K_3 = k_3Q^2 \sim k_3\mathrm{ch}^2(\beta z)$, Eq. (44) contains an additional "orienting" term proportional to $\theta$. If the mean values of the hyperbolic functions and the solution in the form $\theta = \theta_0\cos(\pi z/L)$ are used for estimates, then the effective (measured) elasticity coefficient can be taken in the form

$$k_3^{ef} = k_3 + \left(\frac{\beta L}{\pi}\right)^2\left[\mu + (\mu + D_a)\frac{\mathrm{sh}\beta L - \beta L}{\mathrm{sh}\beta L + \beta L}\right], \qquad \beta = \sqrt{\frac{|A|}{D_\perp}}.$$

or, passing to the definition $K_3 = k_3Q^2 \rightarrow \overline{K}_3 = k_3\overline{Q^2}$, then

$$K_3^{ef} = \overline{K}_3 + \frac{L}{2\pi^2}\beta\left[(2\mu + D_a)\,\mathrm{sh}(\beta L) - D_a\beta L\right]\mathrm{ch}^{-1}\left(\frac{\beta L}{2}\right), \qquad \overline{K}_3 = k_3\overline{Q}^2.$$

The resultant relations redefine the elasticity coefficient and lead to an undesirable dependence on geometry, in the case considered, on the layer thickness $L$. To minimize this dependence, the following inequality should be satisfied:

$$\frac{L}{2\pi^2}\beta\left[(2\mu + D_a)\,\mathrm{sh}(\beta L) - D_a\beta L\right]\mathrm{ch}^{-1}\left(\frac{\beta L}{2}\right) \ll \overline{K}_3. \tag{45}$$

Unique definitions of the new constants $D_a$ and $\mu$ via four Frank coefficients (11) follow from [26]. If their values are substituted into inequality (45), it is not satisfied except for a certain region near the phase transition as $T \rightarrow T^*$ and $A = \alpha(T^* - T) \rightarrow 0$, $\beta \rightarrow 0$. This conclusion cannot be considered as completely proved because the potential $U$ is not considered in the general form. In the nonlocal case based on the concept of a distributed order parameter, these new constants are independent, and inequality (45) can be satisfied. In particular, inequality (45) is always satisfied if $\mu = \mu_1 + \mu_3 = 0$ and $D_a = D_{\|} - D_\perp = 0$. A definite answer to whether this inequality is satisfied or not can be obtained only after measuring the new constants and taking into account all coefficients in the Landau-de Gennes potential. Nevertheless, it should be noted that conditions similar to (45) with a dependence on geometry did not arise in considering B- and S-deformations on the basis of the initial equations without the short-range approximation with the result in the form of Eq. (38).

### C. Linear disclination in the narrow kernel approximation

Let us consider steady deformation of the NLC on the basis of the equations derived in Section **5** in the approximation of locality or a narrow kernel of the ITOP. For simplicity, we assume that all elastic Frank coefficients are identical constants, $K_1=K_2=K_3=K=kQ^2$, and that the newly introduced coefficients (34) are also identical constants, $D = D_\perp = D_{||}$ and $\mu = \mu_1 = -\mu_3$. Apart from simplifying the equations, the additional condition of identical values of the constants allows one to avoid the contradiction associated with the necessity of satisfying Eq. (45). The director is presented in the trigonometric form as

$$\mathbf{r} = (x,\quad y,\quad z),\ \mathbf{n} = (\sin\theta\cos\varphi,\quad \sin\theta\sin\varphi,\quad \cos\theta),\ \theta = \theta(\mathbf{r}),\ \varphi = \varphi(\mathbf{r}).$$

The expression for the free energy density (33) is written as the functional

$$F = \frac{1}{2}\left\{D(\nabla Q)^2 + \mathbf{P}\cdot\nabla Q^2 + kQ^2\left[(\nabla\theta)^2 + \sin^2\theta(\nabla\varphi)^2\right]\right\} + U(Q),$$
$$\mathbf{P} = \mu(\mathbf{n}\nabla\cdot\mathbf{n} + \mathbf{n}\times\nabla\times\mathbf{n}). \tag{46}$$

In the adopted single-constant approximation, the flexoelectric term with the coefficient $\mu$ in the equilibrium equations vanishes because it has the form of surface energy and its effect reduces to the boundary conditions. Let us now pass to cylindrical geometry $\mathbf{r} = (r,\alpha)$ and consider a linear disclination ($\theta = \pi/2$) located along the $z$ axis at the points $r = 0$. The free energy density functional and the equilibrium equations in polar coordinates obtained from Eq. (46) by applying the variation or from Eq. (35) have the form

$$F = \frac{1}{2}\left\{D(\nabla_\perp Q)^2 + \mathbf{P}\cdot\nabla_\perp Q^2 + kQ^2(\nabla_\perp\varphi)^2\right\} + U(Q),$$
$$\begin{cases} \dfrac{1}{r}\dfrac{\partial}{\partial r}\left(r\dfrac{\partial Q}{\partial r}\right) + \dfrac{Q_{\alpha\alpha}}{r^2} - \dfrac{k}{D}Q\left(\varphi_r^2 + \dfrac{\varphi_\alpha^2}{r^2}\right) = \dfrac{U_Q}{D}, \\ \dfrac{1}{r}\dfrac{\partial}{\partial r}\left(Q^2 r\dfrac{\partial\varphi}{\partial r}\right) + \dfrac{1}{r^2}\dfrac{\partial}{\partial\alpha}\left(Q^2\dfrac{\partial\varphi}{\partial\alpha}\right) = 0, \end{cases} \tag{47}$$

where

$$\mathbf{P}(r,\alpha) = \mu\left(\frac{\varphi_\alpha}{r},\quad -\varphi_r\right),\quad \nabla_\perp = \left(\frac{\partial}{\partial r},\quad \frac{1}{r}\frac{\partial}{\partial\alpha}\right).$$

These equations still contain the azimuthal angle $\varphi$ from the Cartesian coordinate system, which is counted from the $x$ axis. In the cylindrical coordinate system, the angle is counted from the $r$ axis. The relationship between these angles is obvious ($\varphi^{Cart} = \varphi^{cylin} + \alpha$) and does not affect the form of the equilibrium equations. The subscripts are used to indicate the corresponding derivatives. We seek for a solution for which $Q = Q(r)$ and $\varphi = \varphi(\alpha)$. After this substitution, the free energy density functional and the system of equilibrium equations transform to

$$F = \frac{1}{2}\left[D\left(\frac{\partial Q}{\partial r}\right)^2 + kQ^2\frac{1}{r^2}\left(\frac{\partial\varphi}{\partial\alpha}\right)^2\right] + \mu Q\frac{\partial Q}{\partial r}\frac{1}{r}\frac{\partial\varphi}{\partial\alpha} + U(Q);$$
$$\begin{cases} \dfrac{1}{r}\dfrac{\partial}{\partial r}\left(r\dfrac{\partial Q}{\partial r}\right) - \dfrac{k}{D}\dfrac{Q}{r^2}\left(\dfrac{\partial\varphi}{\partial\alpha}\right)^2 = \dfrac{U_Q}{D}, \\ \dfrac{\partial^2\varphi}{\partial\alpha^2} = 0. \end{cases} \tag{48}$$

moreover, we can immediately write the solution for $\varphi$ in the form $\varphi = s\alpha + c$, where $s = \pm 1/2$, $\pm 1$, $\pm 3/2$… is the disclination force and $c$ is the constant of integration, $0 < c < \pi$ [1-3]. Substitution of the solution for $\varphi$ into the equation for the order parameter $Q(r)$ yields the second-order ordinary differential equation

$$\frac{1}{r}\frac{d}{dr}\left(r\frac{dQ}{dr}\right)-\frac{k}{D}Q\frac{s^2}{r^2}=\frac{U_Q}{D},$$

which is nonlinear in the general case for an arbitrary value of $U_Q$. Substituting the Landau-de Gennes potential (32) and normalizing the radial components, we can rewrite the equation as

$$\frac{1}{\xi}\frac{d}{d\xi}\left(\xi\frac{dQ}{d\xi}\right)+a^2Q\left(1-\frac{s^2}{\xi^2}\right)=-bQ^2+cQ^3,\qquad Q\big|_{\xi=0,\infty}<\infty,$$
$$\xi=r\left(\frac{|A|}{k}\right)^{1/2},\quad a=\sqrt{\frac{k}{D}},\quad b=\frac{3kB}{|A|D},\quad c=\frac{4kC}{|A|D}. \qquad (49)$$

The boundary conditions for $Q$ are the limited values at the center ($r$ = 0) and at infinity as $\xi \to \infty$. For a quadratic potential, Eq. (49) reduces to the Bessel equation. As an example, Fig. 4 shows the numerical solutions for the particular case.

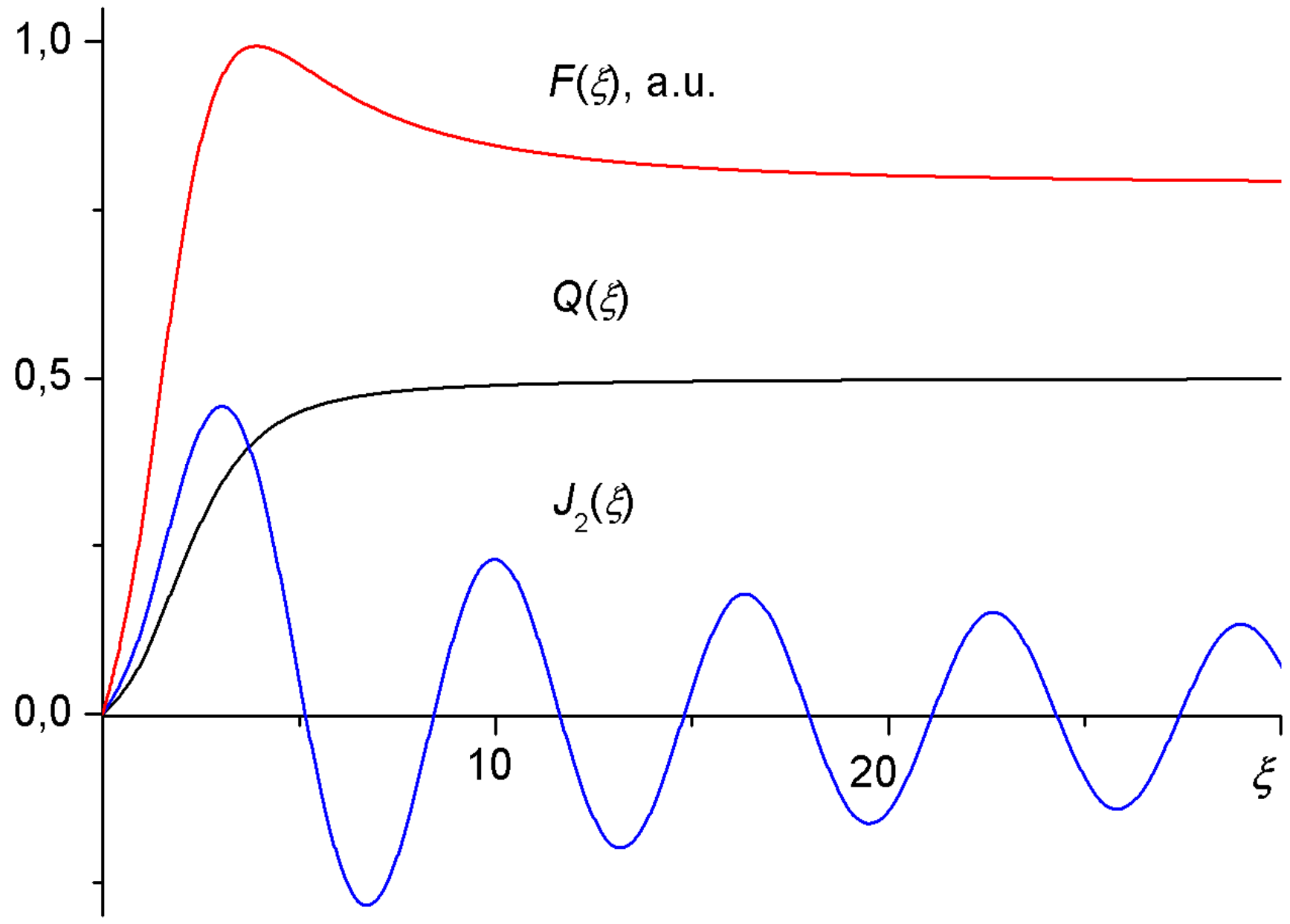

FIG. 4. Scalar order parameter versus the dimensionless radius $\xi$ for the disclination with $s = 2$. The blue curve is the Bessel function $J_2(\xi)$, which is the solution of the linear version of Eq. (49) for $a$ =1 ($k = D$ and $\mu$ = 0). The black curve is the solution of the nonlinear equation (49) for $a$ =1, $b$ = 0, and $c$ = -4. The red curve is the free energy density in arbitrary units for the same parameters.

In the linear approximation or at $U = AQ^2/2$, the solution in the form of the Bessel function is qualitatively different from the general solution. The Bessel functions oscillate with increasing radius, change their sign, and tend to zero (isotropic LC state), which obviously contradicts the experimental observations. For Eqs. (48), the value of $Q$ at infinity ($r \to 0$) is determined from the condition of the zero value of the derivative of the Landau-de Gennes potential,

$$\left.\frac{\partial U}{\partial Q}\right|_{r\to\infty}=Q\left(4CQ^2-3BQ+A\right)=0\to Q\big|_{r\to\infty}=\begin{cases}Q_1=0,\\ Q_{2,3}=\dfrac{3B\pm\sqrt{9B^2-16AC}}{8C}.\end{cases}$$

The roots of this equation are equivalent to similar values obtained in analyzing phase transitions in a nondeformed LC [1-3]. If the director does not coincide with the radial direction of the order parameter gradient, this case can be considered as the emergence of biaxiality induced by deformation; moreover, this effect is obtained within the framework of the initial uniaxial model.

It follows from the solution (Fig. 4) that the LC tends to the isotropic state in the region of the disclination line (formally, this line is also a certain axis) or as $r \to 0$.

Equations similar to (47), with ignored possible influence of the polar vector **P** (at $\mu_1 \neq -\mu_3$), for linear disclinations ($\theta = \pi/2$) were studied in sufficient detail [42, 43] under the assumption that there exists a certain "amplitude" of the director, $\mathbf{N} \sim \sqrt{Q}\mathbf{n}$ (3). The equations were derived on the basis of Ericksen's assumption [18] of introducing terms with scalar order parameter gradients into the free energy. The main conclusion was also the assumption about the possibility of existence of the isotropic phase at the center (on the axis) of the disclination.

Certainly, the particular cases of the new concept of the distributed order parameter do not cover the entire variety of liquid crystals. Nevertheless, the absence of contradictions can be considered as indirect validation of the adequacy of the ITOP concept.

## VII. DISCUSSION OF RESULTS. CONCLUSIONS

The problem of a theoretical basis for nonequilibrium transport processes in various fields of physics has been a topical issue for many years. The development of science and technology requires the assessment of high-velocity fine transitional processes, which are not described by the classical mechanics of continuous media. There arises a need to extend the concept of continuum mechanics to structured media and processes in open systems far from the thermodynamic equilibrium state. Several attempts of this kind were made during the last decades, but all theories have certain drawbacks. The most general results were obtained on a rigorous statistical-mechanical basis; however, it is difficult to use these results because these models are not closed at the macroscopic level. It is principally impossible to close a closed kinetic theory of nonequilibrium processes because of collective interaction effects, which cannot be included into the kinetic equations via the potential. Meanwhile, experimental results obtained in various fields of mechanics (hydrodynamics of turbulent flows, multiphase media, wave processes in solids, etc.) reveal many common features for all these processes. The higher the rates of the processes and the more complicated the internal structure of the medium, the more pronounced these features. They characterize a nonclassical response of the medium to an external perturbation and are manifested as effects of self-organization and self-control, which are always inherent in nonequilibrium transport to a greater or smaller extent. Self-organization as a response of a complex system to an intense external action is the process of formation of multiscale vortex and wave structures, which are not directly related to the primary structure of the system. All these problems refer to the LC description as well.

Some comments concerning other thermodynamic variables should be also made. For thermodynamic consistency, the order parameter should be considered as a certain function of two variables, e.g., density $\rho$ and temperature $T$. For certainty, by virtue of the qualitative character of our discussion, let us specify the scalar order parameter $Q = Q(\rho,T)$ defined in Eq. (22) or introduced in the case of the locality approximation (31). Its spatial changes should be accompanied by changes in density and temperature. The dependence on temperature was extensively studied both theoretically and experimentally [1-3] for weakly deformed LCs. In the isothermal case, structure deformation is expected to induce a change in density, in particular, without this assumption it is impossible to explain the emergence of the isotropic state at the disclination axis (Figs. 3 and 4). The layered structure of SLCs also leads to periodic changes in density, but without deformation. At small values of $\Delta\rho$ and $\Delta T$, it is possible to assume a linear dependence

$$\Delta Q = \frac{\partial Q}{\partial \rho}\Delta\rho + \frac{\partial Q}{\partial T}\Delta T, \tag{50}$$

which implies the inverse possibility in the isothermal case ($\Delta T = 0$): a change in density due to a change in the order parameter. In another variant, where the change in density can be neglected on the background of the change in temperature, the order parameter depends only on temperature. Here we can consider a more general dependence on the scalar order parameter on temperature $Q(T)$ [2] than the liner approximation (50). After substitution of the unique dependence $Q(T)$ into Eqs. (35) and some approximations of no principal significance, it is possible to obtain steady equations derived in [16], where it was done on the basis of other reasoning than here. As was demonstrated in [16], in the presence of the temperature gradient only, the first equation in system (35) should be correlated with the steady equation of heat conduction; in this case, the tensor $D_{ij}$ (33) is linearly related to the heat conduction tensor. Hu and Palffy-Muhoray [17] considered the thermal orientation effect in an LC with a dye induced by the polarizability **P** (33), whose form is similar to the flexoelectric polarizability [1-3]. As was noted above, it is proposed to consider experiments on thermal orientation effects [15, 17] as a certain "tool" for determining new (arising in the present considerations) constants in Eq. (34), $D_{||}$, $D_{\perp}$, $\mu_1$, $\mu_3$, and also for refining the dependence $Q(T)$ itself. The issue of determining the general law for the order parameter $Q = Q(\rho,T)$, including its dependence on density $\rho$, remains open.

We expect that the approach based on constructing nonlocal phenomenological theories for liquid crystals will be successful because a similar approach has been already used to describe many physical phenomena [32, 44, 45]. The main idea of this approach is based on generalization of the Onsager relations [46]:

$$I_n = \sum_k \int L_{nk}(x,x')X_k(x')dx' \quad . \tag{51}$$

Owing to introduction of a new integral tensor order parameter, it is also necessary to generalize other ITOP-dependent physical variables in the Onsager relation (51). For example, obvious relations for the electric induction **D,** heat flux **q,** and diffusion are

$$\begin{aligned} &D_i(\mathbf{r}) = \int \varepsilon_{ij}(\mathbf{r},\mathbf{r}')E_j(\mathbf{r}')dV', \quad \varepsilon_{ij}(\mathbf{r},\mathbf{r}') = \frac{1}{3}\left(\varepsilon'_{\|} + 2\varepsilon'_{\perp}\right)S(\mathbf{r},\mathbf{r}')\delta_{ij} + \varepsilon'_a S_{ij}(\mathbf{r},\mathbf{r}'); \\ &q_i(\mathbf{r}) = \int \kappa_{ij}(\mathbf{r},\mathbf{r}')\frac{\partial T(\mathbf{r}')}{\partial x'_j}dV', \quad \kappa_{ij}(\mathbf{r},\mathbf{r}') = \frac{1}{3}\left(\kappa'_{\|} + 2\kappa'_{\perp}\right)S(\mathbf{r},\mathbf{r}')\delta_{ij} + \kappa'_a S_{ij}(\mathbf{r},\mathbf{r}'); \end{aligned} \tag{52}$$

these relations are written for the uniaxial case (30), and the primed variables are the ITOP-independent dielectric and heat conduction constants along and across the axis. Relaxation coefficients are introduced in a more complicated manner than Eqs. (52), which primarily refers to viscous parameters.

Thus, a new approach to the phenomenological description of the LC state is proposed. This approach is based on the concept of a distributed or integral tensor order parameter. Examples in the present paper demonstrate that this parameter is noncontradictory and consistent with available experimental data. It is also planned to consider other aspects of the proposed theory and to analyze more complicated LC orientation configurations. The proposed statements refer to steady deformed states of liquid crystals and phase transitions in them, but the authors do not consider this fact as a constraint. Later on, it is planned to extend the introduced concepts and computational algorithm developed in the present study to dynamic cases with allowance for the inertia of the moment motion and hydrodynamics of anisotropic fluids under possible application of external fields. The present work is considered as the first stage of solving the problem aimed at the description of small-scale phenomena in LCs, in particular, those mentioned in Introduction. For example, the emergence of radial components of the director and disclination formation [8, 9] in the case of radiation absorption and, correspondingly, heat removal (heat flux) in the radial direction from the line of propagation of the bounded light beam is explained.

To conclude, it is also necessary to perform comparisons with available local theories. In the present paper, we basically refer to publications [26, 37]. Certainly, there are many more works on this topic. For comparisons, however, we only need the general trend of taking into account the entire variety of LC properties, which actually reduces to expanding of the elastic part of the free energy into the Taylor series in order parameter gradients. This formalism always leads to a correct result if a necessary number of terms in the expansion are left. In this sense, we propose a certain general dependence of the LC elasticity, which does not require any expansions for refining. In particular, one of the consequences is the linearity of the LC equilibrium equations for the quadratic form of the Landau-de Gennes potential (17, 19) or (28). An obvious advantage is the inequality of all Frank constants, which is impossible in the quadratic approximation within the framework of locality. The main challenge of the present paper is to justify the adequacy of the introduced integral (distributed) notions; because of the limited volume, it cannot be extended to consider other examples, which demonstrate the advantages of the new formalism. This will be done in subsequent papers. Running ahead, we can tell that many other LC states and phase transitions have been already considered; these examples include the effects of external fields (e.g., smectics of various kinds, blue phases of CLCs, and block LCs), which are adequately described on the basis of the concepts and analytical calculations proposed in this work.

The modern trends in the field of the LC state theory, including molecular dynamics and quantum physics, finally reduce to searching for the relationship between the microscopic and macroscopic (continuum, phenomenologically grounded) concepts of the medium state. It should be emphasized that macroscopic parameters can be exactly measured in experiments. The proposed concept of the distributed order parameter will extend the range of the macroscopic presentation of the LC state and will allow determination of new relationships between the microscopic and macroscopic aspects of liquid crystals.

## APPENDIX

The functional of the deformation part of the free energy density (33) is calculated below. The calculation is performed for each term separately and is marked as corresponding to the coefficients $M_1,\dots,M_8$, $M_q$. In all calculations, the derivatives with respect to the coordinate **r** are first taken, and then the coordinate dependence **r**' ($dV'$) is integrated over the volume $V$. The remaining notations are given in Section 5. In calculations we use the relations

$$\frac{\partial n_i}{\partial x_k}\frac{\partial n_i}{\partial x_k}=(\nabla\cdot\mathbf{n})^2+(\nabla\times\mathbf{n})^2,\qquad (\nabla\times\mathbf{n})^2=(\mathbf{n}\cdot\nabla\times\mathbf{n})^2+(\mathbf{n}\times\nabla\times\mathbf{n})^2,$$

$$\frac{\partial n_i}{\partial x_k}n_k=(\mathbf{n}\cdot\nabla)n_i=-(\mathbf{n}\times\nabla\times\mathbf{n})_i.$$

Let us calculate separately each term in Eq. (32).

$$\underline{M_1}:\qquad S_{ijk}S_{ijk}=\left(\frac{\partial Q}{\partial x_k}\right)^2+\frac{1}{3}\left(\frac{\partial Q}{\partial x_k}\right)^2+\left(Q\frac{\partial n_i}{\partial x_k}\right)^2-\frac{2}{3}\left(\frac{\partial Q}{\partial x_k}\right)^2 n_j n_j'-\frac{2}{3}\frac{\partial Q}{\partial x_k}Q\frac{\partial n_i' n_i}{\partial x_k},$$

$$\int S_{ijk}S_{ijk}\delta(\mathbf{r}-\mathbf{r}')\,dV'=\frac{2}{3}\left(\frac{\partial Q}{\partial x_k}\right)^2+Q^2\left(\frac{\partial n_i}{\partial x_k}\right)^2=\frac{2}{3}(\nabla Q)^2+Q^2\left(\frac{\partial n_i}{\partial x_k}\right)^2.$$

$$\underline{M_2}:\quad S_{ikk}S_{ijj}=\frac{\partial Q}{\partial x_k}n'_j n'_k\frac{\partial Q}{\partial x_j}-\frac{1}{3}\frac{\partial Q}{\partial x_k}n'_k n_i\frac{\partial Q}{\partial x_i}-\frac{1}{3}\frac{\partial Q}{\partial x_i}n_i n'_j\frac{\partial Q}{\partial x_j}+\frac{1}{9}\left(\frac{\partial Q}{\partial x_i}\right)^2-\frac{1}{3}Q\frac{\partial Q}{\partial x_i}n'_j\frac{\partial n_i}{\partial x_j}-$$

$$-\frac{1}{3}Q\frac{\partial Q}{\partial x_i}n'_k\frac{\partial n_i}{\partial x_k}+Q^2\frac{\partial n_i}{\partial x_k}n'_k n'_j\frac{\partial n_i}{\partial x_j},$$

$$\int S_{ikk}S_{ijj}\delta(\mathbf{r}-\mathbf{r}')\,dV'=\frac{1}{3}\frac{\partial Q}{\partial x_k}\left(n_j n_k+\frac{1}{3}\delta_{jk}\right)\frac{\partial Q}{\partial x_j}-\frac{2}{3}Q\frac{\partial Q}{\partial x_i}n_j\frac{\partial n_i}{\partial x_j}+Q^2\frac{\partial n_i}{\partial x_k}n_k n_j\frac{\partial n_i}{\partial x_j}=$$

$$=\frac{1}{3}\frac{\partial Q}{\partial x_k}\left(n_j n_k+\frac{1}{3}\delta_{jk}\right)\frac{\partial Q}{\partial x_j}+\frac{1}{3}\nabla Q^2\cdot(\mathbf{n}\times\nabla\times\mathbf{n})+Q^2(\mathbf{n}\times\nabla\times\mathbf{n})^2.$$

$$\underline{M_3}:\quad S_{ijk}S_{ikj}=\left(n_i n'_j\frac{\partial Q}{\partial x_k}-\frac{1}{3}\delta_{ij}\frac{\partial Q}{\partial x_k}+Qn'_j\frac{\partial n_i}{\partial x_k}\right)\left(n_i n'_k\frac{\partial Q}{\partial x_j}-\frac{1}{3}\delta_{ik}\frac{\partial Q}{\partial x_j}+Qn'_k\frac{\partial n_i}{\partial x_j}\right)=$$

$$=\frac{\partial Q}{\partial x_k}\left(n'_k n'_j-\frac{1}{3}n'_j n_k-\frac{1}{3}n'_k n_j+\frac{\delta_{kj}}{9}\right)\frac{\partial Q}{\partial x_j}-\frac{2}{3}\frac{\partial Q}{\partial x_j}Qn'_j\frac{\partial n_k}{\partial x_k}+Q^2\frac{\partial n_i}{\partial x_k}n'_k n'_j\frac{\partial n_i}{\partial x_j},$$

$$\int S_{ijk}S_{ikj}\delta(\mathbf{r}-\mathbf{r}')\,d\mathbf{r}'=\frac{1}{3}\frac{\partial Q}{\partial x_k}\left(n_k n_j+\frac{\delta_{kj}}{3}\right)\frac{\partial Q}{\partial x_j}-\frac{2}{3}\frac{\partial Q}{\partial x_j}Qn_j\frac{\partial n_k}{\partial x_k}+Q^2\frac{\partial n_i}{\partial x_k}n_k n_j\frac{\partial n_i}{\partial x_j}=$$

$$=\frac{1}{3}\frac{\partial Q}{\partial x_k}\left(n_k n_j+\frac{\delta_{kj}}{3}\right)\frac{\partial Q}{\partial x_j}-\frac{1}{3}(\mathbf{n}\cdot\nabla Q^2)(\nabla\cdot\mathbf{n})+Q^2(\mathbf{n}\times\nabla\times\mathbf{n})^2.$$

$$\underline{M_4}:\quad S_{jik}S_{ijk}=\left(\frac{\partial Q}{\partial x_k}n_j n'_i-\frac{1}{3}\frac{\partial Q}{\partial x_k}\delta_{ij}+Qn'_i\frac{\partial n_j}{\partial x_k}\right)\left(\frac{\partial Q}{\partial x_k}n_i n'_j-\frac{1}{3}\frac{\partial Q}{\partial x_k}\delta_{ij}+Qn'_j\frac{\partial n_i}{\partial x_k}\right)=$$

$$=\frac{1}{3}\left(\frac{\partial Q}{\partial x_k}\right)^2\left(4-n_i n'_i-n_j n'_j\right)+Q\frac{\partial Q}{\partial x_k}\left(2n_j n'_j n'_i\frac{\partial n_i}{\partial x_k}-\frac{1}{3}n'_i\frac{\partial n_i}{\partial x_k}-\frac{1}{3}n'_j\frac{\partial n_j}{\partial x_k}\right)+Q^2\frac{\partial n_j}{\partial x_k}n'_j n'_i\frac{\partial n_i}{\partial x_k},$$

$$\int S_{jik}S_{ijk}\delta(\mathbf{r}-\mathbf{r}')\,dV'=\frac{2}{3}\left(\frac{\partial Q}{\partial x_k}\right)^2=\frac{2}{3}(\nabla Q)^2.$$

$$\underline{M_5}:\quad S_{kik}S_{ijj}=\left(\frac{\partial Q}{\partial x_k}n_k n'_i-\frac{1}{3}\delta_{ik}\frac{\partial Q}{\partial x_k}+Qn'_i\frac{\partial n_k}{\partial x_k}\right)\left(\frac{\partial Q}{\partial x_j}n_i n'_j-\frac{1}{3}\delta_{ij}\frac{\partial Q}{\partial x_j}+Qn'_j\frac{\partial n_i}{\partial x_j}\right)=$$

$$=\frac{\partial Q}{\partial x_k}\left(n_k n'_i n_i n'_j-\frac{2}{3}n_k n'_j+\frac{1}{9}\delta_{kj}\right)\frac{\partial Q}{\partial x_j}+$$

$$+Q\frac{\partial Q}{\partial x_k}\left(n_k n'_i-\frac{1}{3}\delta_{ik}\right)n'_j\frac{\partial n_i}{\partial x_j}+Q\frac{\partial Q}{\partial x_j}n'_j\left(n_i n'_i-\frac{1}{3}\right)\frac{\partial n_k}{\partial x_k}+Q^2\frac{\partial n_k}{\partial x_k}n'_i n'_j\frac{\partial n_i}{\partial x_j},$$

$$\int S_{kik}S_{ijj}\delta(\mathbf{r}-\mathbf{r}')\,dV'=\frac{1}{3}\frac{\partial Q}{\partial x_k}\left(n_k n_j+\frac{1}{3}\delta_{kj}\right)\frac{\partial Q}{\partial x_j}-\frac{1}{3}Q\frac{\partial Q}{\partial x_k}n_j\frac{\partial n_k}{\partial x_j}+\frac{2}{3}Q\frac{\partial Q}{\partial x_j}n_j\frac{\partial n_k}{\partial x_k}=$$

$$=\frac{1}{3}\frac{\partial Q}{\partial x_k}\left(n_k n_j+\frac{1}{3}\delta_{kj}\right)\frac{\partial Q}{\partial x_j}+\frac{1}{3}Q\nabla Q\cdot(\mathbf{n}\times\nabla\times\mathbf{n})+\frac{2}{3}Q(\mathbf{n}\cdot\nabla Q)(\nabla\cdot\mathbf{n}).$$

$$\underline{M_6}: \quad S_{jik}S_{ikj} = \left(\frac{\partial Q}{\partial x_k} n_j n_i' - \frac{1}{3}\delta_{ij}\frac{\partial Q}{\partial x_k} + Qn_i'\frac{\partial n_j}{\partial x_k}\right)\left(\frac{\partial Q}{\partial x_j} n_i n_k' - \frac{1}{3}\delta_{ik}\frac{\partial Q}{\partial x_j} + Qn_k'\frac{\partial n_i}{\partial x_j}\right) =$$

$$= \frac{\partial Q}{\partial x_k}\left(n_j n_i' n_i n_k' - \frac{1}{3}n_k' n_j - \frac{1}{3}n_j n_k' + \frac{\delta_{jk}}{9}\right)\frac{\partial Q}{\partial x_j} + Q\frac{\partial Q}{\partial x_k} n_j n_i' n_k' \frac{\partial n_i}{\partial x_j} + Q\frac{\partial Q}{\partial x_j}\frac{\partial n_j}{\partial x_k} n_i' n_i n_k' -$$

$$-\frac{1}{3}Q\frac{\partial Q}{\partial x_k} n_k' \frac{\partial n_i}{\partial x_i} - \frac{1}{3}Q\frac{\partial Q}{\partial x_j} n_k' \frac{\partial n_j}{\partial x_k} + Q^2\frac{\partial n_j}{\partial x_k} n_i' n_k' \frac{\partial n_i}{\partial x_j},$$

$$\int S_{jik}S_{ikj}\delta(\mathbf{r}-\mathbf{r}')\,dV' = \frac{1}{3}\frac{\partial Q}{\partial x_k}\left(n_j n_k + \frac{\delta_{jk}}{3}\right)\frac{\partial Q}{\partial x_j} + \frac{2}{3}Q\frac{\partial Q}{\partial x_j}\frac{\partial n_j}{\partial x_k} n_k - \frac{1}{3}Q\frac{\partial Q}{\partial x_k} n_k \frac{\partial n_i}{\partial x_i} =$$

$$= \frac{1}{3}\frac{\partial Q}{\partial x_k}\left(n_j n_k + \frac{\delta_{jk}}{3}\right)\frac{\partial Q}{\partial x_j} - \frac{2}{3}Q\nabla Q\cdot(\mathbf{n}\times\nabla\times\mathbf{n}) - \frac{1}{3}Q(\mathbf{n}\cdot\nabla Q)(\nabla\cdot\mathbf{n}).$$

$$\underline{M_7}: \quad S_{kik}S_{jij} = \left(\frac{\partial Q}{\partial x_k} n_k n_i' - \frac{1}{3}\delta_{ki}\frac{\partial Q}{\partial x_k} + Qn_i'\frac{\partial n_k}{\partial x_k}\right)\left(\frac{\partial Q}{\partial x_j} n_j n_i' - \frac{1}{3}\delta_{ji}\frac{\partial Q}{\partial x_j} + Qn_i'\frac{\partial n_j}{\partial x_j}\right) =$$

$$= \frac{\partial Q}{\partial x_k} n_k n_j \frac{\partial Q}{\partial x_j} - \frac{1}{3}\frac{\partial Q}{\partial x_j}\frac{\partial Q}{\partial x_k} n_k n_j' + Q\frac{\partial Q}{\partial x_k} n_k \frac{\partial n_j}{\partial x_j} - \frac{1}{3}\frac{\partial Q}{\partial x_k}\frac{\partial Q}{\partial x_j} n_j n_k' + \frac{1}{9}\frac{\partial Q}{\partial x_k}\frac{\partial Q}{\partial x_k} -$$

$$-\frac{1}{3}Q\frac{\partial Q}{\partial x_k} n_k' \frac{\partial n_j}{\partial x_j} + \frac{\partial n_k}{\partial x_k} Q\frac{\partial Q}{\partial x_j} n_j - \frac{1}{3}Q\frac{\partial Q}{\partial x_j} n_j' \frac{\partial n_k}{\partial x_k} + Q^2\frac{\partial n_k}{\partial x_k}\frac{\partial n_j}{\partial x_j},$$

$$\int S_{kik}S_{jij}\delta(\mathbf{r}-\mathbf{r}')\,dV' = \frac{1}{3}\frac{\partial Q}{\partial x_k}\left(n_k n_j + \frac{\delta_{jk}}{3}\right)\frac{\partial Q}{\partial x_j} + \frac{4}{3}Q\frac{\partial Q}{\partial x_k} n_k \frac{\partial n_j}{\partial x_j} + Q^2\frac{\partial n_k}{\partial x_k}\frac{\partial n_j}{\partial x_j} =$$

$$= \frac{1}{3}\frac{\partial Q}{\partial x_k}\left(n_k n_j + \frac{\delta_{jk}}{3}\right)\frac{\partial Q}{\partial x_j} + \frac{4}{3}Q(\mathbf{n}\cdot\nabla Q)(\nabla\cdot\mathbf{n}) + Q^2(\nabla\cdot\mathbf{n})^2.$$

$$\underline{M_8}: \quad S_{jik}S_{kij} = \left(\frac{\partial Q}{\partial x_k} n_j n_i' - \frac{1}{3}\frac{\partial Q}{\partial x_k}\delta_{ij} + Qn_i'\frac{\partial n_j}{\partial x_k}\right)\left(\frac{\partial Q}{\partial x_j} n_k n_i' - \frac{1}{3}\frac{\partial Q}{\partial x_j}\delta_{ik} + Qn_i'\frac{\partial n_k}{\partial x_j}\right) =$$

$$= \frac{\partial Q}{\partial x_k}\left(n_j n_k - \frac{1}{3}n_k' n_j - \frac{1}{3}n_k n_j' + \frac{\delta_{jk}}{9}\right)\frac{\partial Q}{\partial x_j} + Q\frac{\partial Q}{\partial x_k}\left(n_j - \frac{1}{3}n_j'\right)\frac{\partial n_k}{\partial x_j} + Q\frac{\partial Q}{\partial x_j}\left(n_k - \frac{1}{3}n_k'\right)\frac{\partial n_j}{\partial x_k} +$$

$$+Q^2\frac{\partial n_j}{\partial x_k}\frac{\partial n_k}{\partial x_j} = \frac{\partial Q}{\partial x_k}\left(n_j n_k - \frac{1}{3}n_k' n_j - \frac{1}{3}n_k n_j' + \frac{\delta_{jk}}{9}\right)\frac{\partial Q}{\partial x_j} + 2Q\frac{\partial Q}{\partial x_k}\left(n_j - \frac{1}{3}n_j'\right)\frac{\partial n_k}{\partial x_j} + Q^2\frac{\partial n_j}{\partial x_k}\frac{\partial n_k}{\partial x_j},$$

$$\int S_{jik}S_{kij}\delta(\mathbf{r}-\mathbf{r}')\,dV' = \frac{1}{3}\frac{\partial S}{\partial x_k}\left(n_j n_k + \frac{\delta_{jk}}{3}\right)\frac{\partial S}{\partial x_j} + \frac{4}{3}S\frac{\partial S}{\partial x_k} n_j \frac{\partial n_k}{\partial x_j} + S^2\frac{\partial n_j}{\partial x_k}\frac{\partial n_k}{\partial x_j} =$$

$$= \frac{1}{3}\frac{\partial S}{\partial x_k}\left(n_j n_k + \frac{\delta_{jk}}{3}\right)\frac{\partial S}{\partial x_j} - \frac{4}{3}S\nabla S\cdot(\mathbf{n}\times\nabla\times\mathbf{n}) + S^2\frac{\partial n_j}{\partial x_k}\frac{\partial n_k}{\partial x_j}.$$

$$\underline{M_q}: \quad \varepsilon_{ijk}S_{il}S_{klj} = Q\varepsilon_{ijk}\left(n_i n_l' - \frac{\delta_{il}}{3}\right)\left(\frac{\partial Q}{\partial x_j} n_k n_l' - \frac{1}{3}\delta_{kl}\frac{\partial Q}{\partial x_j} + Qn_l'\frac{\partial n_k}{\partial x_j}\right) =$$

$$= Q\varepsilon_{ijk}\left(n_i\frac{\partial Q}{\partial x_j} n_k n_l' n_l' - \frac{1}{3}\frac{\partial Q}{\partial x_j} n_i\delta_{kl} n_l' + Qn_i n_l' n_l' \frac{\partial n_k}{\partial x_j} - \frac{1}{3}\frac{\partial Q}{\partial x_j} n_k \delta_{il} n_l' + \frac{1}{9}\frac{\partial Q}{\partial x_j}\delta_{kl}\delta_{il} - \frac{1}{3}Q\delta_{il} n_l' \frac{\partial n_k}{\partial x_j}\right),$$

$$\varepsilon_{ijk}\int S_{il}S_{klj}\delta(\mathbf{r}-\mathbf{r}')\,dV' = \frac{2}{3}Q^2 n_i \varepsilon_{ijk}\frac{\partial n_k}{\partial x_j} = \frac{2}{3}Q^2\mathbf{n}\cdot\nabla\times\mathbf{n}.$$

**ACKNOWLEDGMENTS**

The authors would like to express their sincere gratitude to Prof. S.K. Godunov for numerous discussions on various issues of thermodynamic consistency of natural phenomena, including liquid crystals and other soft media.

---